\documentclass{ps}
\usepackage{amsmath, amssymb, amsthm}
\usepackage{graphicx}
\usepackage{bm}
\usepackage{upgreek}
\usepackage{enumitem}
\usepackage[ruled]{algorithm}
\usepackage{algorithmic}
\usepackage{hyperref}
\hypersetup{colorlinks,linkcolor={blue},citecolor={blue},urlcolor={blue}}
\usepackage{cleveref}
\usepackage{xcolor}
\usepackage{wrapfig}
\usepackage[a4paper, margin=1in]{geometry}

\newcommand{\norm}[1]{\ensuremath{\Vert #1 \Vert}}

\DeclareMathOperator{\divergence}{div}

\Crefname{assumptionH}{\textbf{H}\hspace{-3pt}}{\textbf{H}\hspace{-3pt}}
\crefname{assumptionH}{\textbf{H}}{\textbf{H}}

\newtheorem{assumptionA}{\textbf{A}\hspace{-3pt}}
\Crefname{assumptionA}{\textbf{A}\hspace{-3pt}}{\textbf{H}\hspace{-3pt}}
\crefname{assumptionA}{\textbf{A}}{\textbf{A}}

\newtheorem{assumptionB}{\textbf{B}\hspace{-3pt}}
\Crefname{assumptionB}{\textbf{B}\hspace{-3pt}}{\textbf{H}\hspace{-3pt}}
\crefname{assumptionB}{\textbf{B}}{\textbf{B}}

\newtheorem{example}{Example}
\begin{document}
\title{Interacting Particle Langevin Algorithm for Maximum Marginal Likelihood Estimation}
\author{O. Deniz Akyildiz}\address{Department of Mathematics, Imperial College London, UK. \texttt{deniz.akyildiz@imperial.ac.uk}} 
\author{Francesca Romana Crucinio}\address{Department of Economics, Social Studies, Applied Mathematics and Statistics, University of Turin, Italy. \\ \texttt{francescaromana.crucinio@unito.it}} 
\author{Mark Girolami}\address{Department of Engineering, University of Cambridge and The Alan Turing Institute, UK. \texttt{mag92@cam.ac.uk}}
\author{Tim Johnston}\address{Universite Paris Dauphine, Paris, France. \texttt{T.Johnston-4@sms.ed.ac.uk}}
\author{Sotirios Sabanis}\address{School of Mathematics, University of Edinburgh, UK; The Alan Turing Institute, UK; National University of Athens, Greece. \texttt{s.sabanis@ed.ac.uk}}

%
%
\begin{abstract} We develop a class of interacting particle systems for implementing a maximum marginal likelihood estimation (MMLE) procedure to estimate the parameters of a latent variable model. We achieve this by formulating a continuous-time interacting particle system which can be seen as a Langevin diffusion over an extended state space of parameters and latent variables. In particular, we prove that the parameter marginal of the stationary measure of this diffusion has the form of a Gibbs measure where number of particles acts as \textit{the inverse temperature} parameter in classical settings for global optimisation. Using a particular rescaling, we then prove geometric ergodicity of this system and bound the discretisation error in a manner that is uniform in time and does not increase with the number of particles. The discretisation results in an algorithm, termed \textit{Interacting Particle Langevin Algorithm} (IPLA) which can be used for MMLE. We further prove nonasymptotic bounds for the optimisation error of our estimator in terms of key parameters of the problem, and also extend this result to the case of stochastic gradients covering practical scenarios. We provide numerical experiments to illustrate the empirical behaviour of our algorithm in the context of logistic regression with verifiable assumptions. Our setting provides a straightforward way to implement a diffusion-based optimisation routine compared to more classical approaches such as the Expectation Maximisation (EM) algorithm, and allows for especially explicit nonasymptotic bounds. \end{abstract}
%
%
\maketitle
\section{Introduction}
Parameter estimation in the presence of hidden, latent, or unobserved variables is key in modern statistical practice. This setting arises in
multiple applications such as frequentist inference \cite{casella2021statistical},  empirical Bayes \cite{carlin2000empirical} and most notably latent variable models to model
complex datasets such as images \cite{bishop2006pattern}, text \cite{blei2003latent}, audio \cite{smaragdis2006probabilistic}, and graphs \cite{hoff2002latent}.
Latent variable models (LVMs) allow us to handle complex datasets by decomposing them into a set of trainable parameters and a set of latent
variables which correspond to the underlying structure of the data. However, learning models with hidden, unobserved, or latent variables is challenging as, in order to find a maximum likelihood estimate (MLE), we first need to tackle an intractable integral.

Broadly speaking, the problem of parameter inference in models with latent variables
 can be formalized as a maximum marginal likelihood estimation problem \cite{dempster1977maximum}.
In this setting, there are three main ingredients of the probabilistic model: The parameters, the latent variables, and the observed
data. More precisely, in maximum marginal likelihood estimation we consider a probabilistic model with observed data $y$ and likelihood
$p_\theta(x, y)$ parametrised by $\theta\in \mathbb{R}^{d_\theta}$, where $x\in\mathbb{R}^{d_x}$ is a latent variable which cannot be observed.
Our aim is to find $\bar{\theta}^\star$ maximising the marginal likelihood $p_\theta(y):=\int_{\mathbb{R}^{d_x}} p_\theta(x, y)d x$. In the literature, the
standard way to solve maximum marginal likelihood estimation problems is to use the Expectation Maximisation (EM)
algorithm \cite{dempster1977maximum}, an iterative procedure which is guarantee to converge to a (potentially local) maximiser of $p_\theta(y)$. The EM algorithm consists of two steps, where the first step is an expectation step (\textit{E-step}) w.r.t. the \textit{latent variables}
 while the second step is a maximisation step (\textit{M-step}) w.r.t. the  \textit{parameters}. While initially this algorithm was mainly implemented in
 the case where both steps are exact, it has been shown that the EM algorithm can be implemented with approximate steps, e.g., using Monte Carlo methods for
 the E-step \cite{wei1990monte, sherman1999conditions}. Similarly, the M-step can be approximated using
 numerical optimisation algorithms \cite{meng1993maximum, liu1994ecme}, e.g., gradient descent \cite{lange1995gradient}. The ubiquity of latent variable models and the effectiveness of
 the EM algorithm compared to its alternatives has led to a flurry of research in the last decades for its approximate implementation. In particular,
 a class of algorithms called Monte Carlo EM (MCEM; \cite{wei1990monte}) or stochastic EM (SEM; \cite{celeux1985sem}) algorithms  have been extensively analysed and used in the literature, see, e.g.,
  \cite{celeux1992stochastic, chan1995monte, sherman1999conditions, booth1999maximizing, cappe1999simulation, diebolt1995stochastic}. This method replaces the E-step with Monte Carlo integration.
  However, since the expectation in the E-step is w.r.t. the posterior distribution of the latent variables given the data, the perfect Monte Carlo integration
  is often intractable. As a result, a plethora of approximations using Markov chain Monte Carlo (MCMC) for the E-step have been proposed and the convergence properties
  of the resulting algorithms have been studied, see, e.g., \cite{delyon1999convergence, fort2011convergence, fort2003convergence, caffo2005ascent, atchade2017perturbed}.
  Standard Markov chain Monte Carlo (MCMC) methods, however, may be hard to implement and slow to converge, particularly when the dimension of the latent
variables is high. This is due to the fact that standard MCMC (Metropolis-based) kernels may be hard to calibrate and may get stuck in local modes easily.
To alleviate such problems, a number of \textit{unadjusted} Markov chain Monte Carlo methods have been proposed, most notably, the unadjusted Langevin algorithm (ULA)
 \cite{durmus2017nonasymptotic,  dalalyan2017further, dalalyan2017theoretical} based on Langevin diffusions \cite{roberts1996exponential}. These approaches are based on a simple discretisation
 of the Langevin stochastic differential equation (SDE) and have become a popular choice for MCMC methods in high dimensional setting with
 favourable theoretical properties, see, e.g., \cite{durmus2019high, vempala2019rapid, durmus2019analysis, brosse2019tamed, dalalyan2019user, chewi2022analysis} and references therein.

Naturally, MCMC methods based on unadjusted Markov kernels like ULA have been also used in the context of EM. In particular, \cite{de2021efficient} studied the SOUL algorithm
which uses ULA (or, more generally, inexact Markov kernels) in order to draw (approximate) samples from the posterior distribution of the latent variables and approximate the E-step.
This algorithm proceeds in a \textit{coordinate-wise} manner, first running a sequence of ULA steps to obtain samples, then approximating the E-step with these samples, and finally moving to the M-step. The convergence analysis has been done under convexity and mixing assumptions but the proofs are based on a particular selection of step-sizes which obfuscates the general applicability of the results. More recently, \cite{kuntz2023particle} showed that interacting Langevin dynamics can substantially improve the performance.
In particular, \cite{kuntz2023particle} used an \textit{interacting} particle system to approximate the E-step, rather than running a ULA chain. This approach has significantly
improved empirical results, however, \cite{kuntz2023particle} provided a limited theoretical analysis. The recent work \cite{caprio2024error}, which appeared after the completion (and arXiv submission) of this work, provides a detailed analysis of the error of the interacting particle system introduced in \cite{kuntz2023particle}. Their results agree with ours, however the proof techniques are significantly different.

Looking at marginal maximum likelihood estimation from an optimisation point of view, \cite{gaetan2003multiple, jacquier2007mcmc, johansen2008particle} propose strategies based on simulated annealing to approximate $\bar{\theta}^\star$.
In particular, \cite{johansen2008particle, duan2017maximum} consider interacting particle methods based on sequential Monte Carlo. In the optimisation literature algorithms based on interacting particle systems are often used in place of gradient based methods
(e.g. \cite{pinnau2017consensus, kennedy1995particle, totzeck2020consensus, akyildiz2020parallel}, see also \cite{grassi2021particle} for a recent review)
or to speed up convergence in gradient-based methods (e.g. \cite{borovykh2021optimizing}).
However, to the best of our knowledge, the use of interacting particle systems combined with Langevin dynamics in the context of maximum
likelihood estimation has not been widely explored,
except for the recent developments in \cite{kuntz2023particle} which paved the way for the present work. \\

\noindent\textbf{Contributions.} In this paper, we propose and analyse a new algorithm for maximum marginal likelihood estimation in the context of latent variable models. Our algorithm forms an
interacting particle system (IPS) which is based on a time discretisation of a continuous time IPS to approximate the maximum marginal likelihood estimator for models with
incomplete or hidden data. Our method uses a similar continuous time system to that of \cite{kuntz2023particle} but with a crucial difference in the parameter $(\theta)$ dimension. Specifically, similar to
\cite{kuntz2023particle}, our IPS model is defined over a set of particles $\{X_1, \ldots, X_N\}$, where $X_i\in \mathbb{R}^{d_x}$, and a parameter $\theta\in \mathbb{R}^{d_\theta}$. However, we inject noise into the dynamics of $\theta$, hence obtain an SDE for $\theta$ rather than a deterministic ordinary differential equation (ODE). This allows us to obtain a system with the invariant measure of the form
\begin{equation}
\label{eq:invariant_measure}
\pi^N_\star(\theta, x_1, ..., x_N) \propto e^{-\sum_{i=1}^N U(\theta, x_i)}
\end{equation}
where $U(\theta, x) := -\log p_\theta(x, y)$ with fixed data $y$. We prove that the $\theta$-marginal of this density concentrates around $\bar{\theta}^\star$ as $N \to \infty$ allowing us to obtain
nonasymptotic results for the algorithm.
It follows that any method of sampling from $\pi^N_\star$ (keeping only the $\theta$-component) will be a method for maximum-marginal likelihood estimation. In fact, samplers targeting $\pi^N_\star$ have been developed both in the MCMC \cite{gaetan2003multiple, jacquier2007mcmc, doucet2002marginal} and in the sequential Monte Carlo literature \cite{johansen2008particle}.

Our approach leverages the rich connection between
Langevin dynamics and optimisation (see, e.g., \cite{dalalyan2017further, pmlr-v65-raginsky17a, zhang2023nonasymptotic}) and opens up further avenues for proving new results. Using a particular natural rescaling, we are additionally able to prove ergodic bounds and numerical error bounds that are completely independent of the parameter $N$. As the first nonasymptotic results under the
interacting setting, our results are obtained in the case where $U$ is gradient Lipschitz and obeys a strong convexity condition,
but we believe that similar results can be obtained under much weaker (nonconvex) conditions \cite{zhang2023nonasymptotic}. In this setting the addition of noise into the $\theta$-component will be crucial,  however we leave this direction for future work.

The remainder of this paper is organised as follows. In Section~\ref{sec:optimisation}, we set up our framework and summarise the key contributions, the proposed particle system and resulting algorithm, and explain in detail the intuition for our main result.  In Section~\ref{sec: analysis}, we prove our results with full proofs and provide global error bounds. Section~\ref{sec: stochastic gradient} extends our results to the case in which the gradients of the potential $U$ are approximated by unbiased estimators. Proofs will be postponed to Appendix~\ref{app:ergodicity}--\ref{app:gf}. In Section~\ref{sec:related} we review closely related work, making our contributions and main ideas clearer. We then provide in Section~\ref{sec:experiments} an example on logistic regression which obeys our assumptions. Finally, we conclude with Section~\ref{sec:conclusion}.

\subsection{Notation}

We endow $\mathbb{R}^d$ with the Borel $\sigma$-field $\mathcal{B}(\mathbb{R}^d)$ with respect to
the Euclidean norm $\norm{\cdot}$. 
For $\alpha \in \mathbb{N}\cup\{ \infty\}$ we let $C^\alpha(\mathbb{R}^d)$ denote the space of $\alpha$-times differentiable functions on $\mathbb{R}^d$, and let $C_c^\alpha(\mathbb{R}^d)\subset C^\alpha(\mathbb{R}^d)$ the subspace of $C^\alpha(\mathbb{R}^d)$ consisting of functions with compact support.
For all continuously differentiable functions $f$,
we denote by $\nabla f$ the gradient. Furthermore, if $f$ is twice continuously differentiable we
denote by $\nabla^2f$ its Hessian and by $\Delta f$ its Laplacian.
For any $p \in \mathbb{N}$ we denote by
$\mathcal{P}_p(\mathbb{R}^d) = \{\pi \in \mathcal{P}(\mathbb{R}^d):\int_{\mathbb{R}^d}
  \norm{x}_p^p d \pi(x) < +\infty\}$ the set of probability measures over
$\mathcal{B}(\mathbb{R}^d)$ with finite $p$-th moment. For ease of notation, we define
$\mathcal{P}(\mathbb{R}^d) = \mathcal{P}_0(\mathbb{R}^d)$ the set of probability measures over
$\mathcal{B}(\mathbb{R}^d)$ and endow this space with the topology of weak convergence.  For any $\mu,\nu\in\mathcal{P}_p(\mathbb{R}^d)$ we define the
$p$-Wasserstein distance $W_p(\mu, \nu)$ between $\mu$ and $\nu$ by
\begin{equation*}
    W_p(\mu, \nu) =\left( \inf_{\gamma \in \mathbf{T}(\mu,\nu)} \int_{\mathbb{R}^d\times \mathbb{R}^d}  \norm{x -y}_p^p d \gamma (x,y)\right)^{1/p}
\end{equation*}
where
$\mathbf{T}(\mu, \nu)=\{\gamma\in\mathcal{P}(\mathbb{R}^d\times
  \mathbb{R}^d):\gamma(A \times \mathbb{R}^d) = \mu(A),\ \gamma(\mathbb{R}^d \times
  A) = \nu(A)\ \forall A \in \mathcal{B}(\mathbb{R}^d)\}$ denotes the set of all couplings between $\mu$ and $\nu$. In the following, we metrise $\mathcal{P}_p(\mathbb{R}^d)$ with $W_p$.

In the following we adopt the convention that the solution to the continuous time solution is given by bold letters, whilst other processes (including the time discretisation) are not.

\section{Background and main results}
\label{sec:optimisation}
In this section, we introduce the problem and technical background of relevant work.
\subsection{Maximum marginal likelihood estimation and the EM algorithm}

In this section, we introduce the problem of maximum marginal likelihood estimation in models with latent, hidden, or unobserved variables. 

Let $p_\theta(x, \cdot)$ be a joint probability density function of the latent variable $x$ and fixed observed data. We define the negative log-likelihood as
\begin{equation*}
    U(\theta, x) := - \log p_\theta(x, y)
\end{equation*}
for any fixed $y\in \mathbb{R}^{d_y}$.
The goal of maximum marginal likelihood estimation (MMLE) is to find the parameter $\bar{\theta}^\star$ that maximises the marginal likelihood \cite{dempster1977maximum}
\begin{equation}\label{eq:def:k}
k(\theta) := p_\theta(y) = \int p_\theta(x, y) d x = \int e^{-U(\theta, x)}d x,
\end{equation}
where $y \in \mathbb{R}^{d_y}$ is the observed (fixed) data. Here $p_\theta(y)$ can be interpreted as a function of $\theta$ for fixed observation $y$ as standard in Bayesian literature. Therefore, the problem of maximum marginal likelihood estimation can be seen as maximisation of an intractable integral \eqref{eq:def:k}.

\subsection{A Particle Langevin SDE and its discretisation}
While optimising a function in a standard setup can be done with standard stochastic optimisation techniques in the convex case or with Langevin diffusions in the nonconvex case \cite{zhang2023nonasymptotic}, when the function to be optimised is an intractable integral as in \eqref{eq:def:k}, the problem becomes highly non-trivial. This stems from the fact that any gradient estimate w.r.t. $\theta$ will require sampling $x$ variables to estimate the integral. But since sampling from the unnormalised measure $e^{-U(\theta, x)}$ for fixed $\theta$ in this setting is typically intractable, the samples drawn using numerical schemes would be approximate which would incur bias on the gradient estimates. This is the precise problem investigated in prior works, see, e.g., \cite{atchade2017perturbed} or \cite{de2021efficient} for using MCMC chains to sample from $e^{-U(\theta, x)}$ with fixed $\theta$ and use these samples to estimate the gradient of $k(\theta)$. This approach requires non-trivial assumptions on step-sizes of optimisation schemes and sometimes not even computationally tractable as the sample sizes used to estimate gradients may need to increase over iterations.

A different approach was considered in \cite{kuntz2023particle} which used a \textit{particle system}. In other words, at iteration $n$, \cite{kuntz2023particle} proposed to use $N$ particles $X^{i, N}_n$ to estimate the gradient of $k(\theta)$ which is then minimised using a gradient step. Inspired by this approach, in this work, we propose an interacting Langevin SDE
\begin{align}
    d \bm{\theta}^N_t &= -\frac{1}{N}\sum_{j=1}^N \nabla_{\theta} U(\bm{\theta}^N_t, \bm{X}_t^{j, N})d t+ \sqrt {\frac{2}{N}}d \bm{B}_t^{0,N}, 
 \label{eq:ContIPS_theta} \\
    d \bm{X}_t^{i, N} &= -\nabla_x U(\bm{\theta}^N_t, \bm{X}_t^{i, N})d t + \sqrt{2}d \bm{B}_t^{i, N},\label{eq:ContIPS_x}
\end{align}
for $i=1, \dots, N$, where $\{(\bm{B}_t^{i,N})_{t \geq 0}\}_{i =0}^N$ is a family of independent
Brownian motions. This interacting particle system (IPS) is closely related to the one introduced in \cite{kuntz2023particle}, with one major difference in Eq.~\eqref{eq:ContIPS_theta} where we add the noise term $\sqrt{2/N} d \bm{B}_t^{0,N}$. Despite seemingly small, the addition of the noise term in the $\theta$-component unlocks the way for proving precise nonasymptotic results as we will summarise in Section~\ref{sec:main_results} after introducing our algorithm.

The IPS given in Eqs.~\eqref{eq:ContIPS_theta}--\eqref{eq:ContIPS_x} can be discretised in a number of ways. In this work, we consider a simple Euler-Maruyama discretisation. This can be defined, given $(\theta_0, X_0^{1:N}) \in \mathbb{R}^{d_\theta}\times(\mathbb{R}^d)^{\otimes N}$ and for any $n \in \mathbb{N}$, as
\begin{align}
    \theta_{n+1} &= \theta_n - \frac{\gamma}{N}\sum_{j=1}^N \nabla_{\theta} U(\theta_n, X_n^{j, N}) + \sqrt {\frac{2\gamma}{N}}\xi_{n+1}^{0,N},  \label{eq:IPS_disc_theta}\\
    X_{n+1}^{i, N} &= X_n^{i, N} - \gamma \nabla_x U(\theta_n, X_n^{i, N}) + \sqrt{2\gamma} \xi_{n+1}^{i, N},\label{eq:IPS_disc_x}
\end{align}
where $\gamma>0$ is a time discretisation parameter and $\xi_n^{i,N}:= \bm{B}^{i,N}_{n \gamma}-\bm{B}^{i,N}_{(n-1)\gamma}$ are iid standard Gaussians for $i=0, \dots, N$.
Algorithm~\ref{algo:ipsem} summarises our method to approximate $\bar{\theta}^\star$. We call our resulting method Interacting Particle Langevin Algorithm (IPLA).

%
For future work, we imagine that the addition of noise will allow for our algorithm to be especially effective in the nonconvex case, since this will allow the theta component to escape more easily from shallow local minima. Furthermore, this method could be adapted to an effective annealing algorithm, where more particles are introduced as a function of time. However for the present work we consider the simplest possible case.

\begin{algorithm}
\caption{Interacting Particle Langevin Algorithm (IPLA)}
\label{algo:ipsem}
\begin{algorithmic}
\STATE{\textbf{Require:} $N, \gamma, \pi_{\mathrm{init}} \in \mathcal{P}(\mathbb{R}^{d_\theta})\times\mathcal{P}((\mathbb{R}^d)^N)$}
\STATE{Draw $(\theta_0, \{X_0^{i,N}\}_{i=1}^N)$ from $\pi_{\mathrm{init}}$}
\FOR{$n=1:n_T$}
\STATE{$\theta_{n+1} = \theta_n - \frac{\gamma}{N}\sum_{j=1}^N \nabla_{\theta} U(\theta_n, X_n^{j, N}) + \sqrt {\frac{2\gamma}{N}}\xi_{n+1}^{0,N}$ and \\
    $X_{n+1}^{i, N} = X_n^{i, N} - \gamma \nabla_x U(\theta_n, X_n^{i, N}) + \sqrt{2\gamma} \xi_{n+1}^{i, N}$}
\ENDFOR
\RETURN $\theta_{n_T + 1}$
\end{algorithmic}
\end{algorithm}

\subsection{The proof strategy and the main result}\label{sec:main_results}
In this section, we briefly summarise the main ideas behind our analysis of the scheme given in \eqref{eq:IPS_disc_theta}--\eqref{eq:IPS_disc_x} and our final result which quantifies the convergence of the sequence $(\theta_n)_{n\geq 1}$ to the minimiser of $k$, namely $\bar{\theta}^\star$. A standard result from the literature shows that finding minimisers of $k$ also solves the problem of sampling from the posterior distribution $p_{\bar{\theta}^\star}(x|y)$, see, e.g., \cite{neal1998view} or \cite[Theorem~1]{kuntz2023particle}.

In what follows, we will sketch our results on the invariant measure, the convergence of the IPS \eqref{eq:ContIPS_theta}--\eqref{eq:ContIPS_x} and the discrete algorithm \eqref{eq:IPS_disc_theta}--\eqref{eq:IPS_disc_x}. These results will extend proofs from a standard Langevin diffusion to the much more complicated particle systems we consider. A key aspect of our theoretical framework is that we consider the rescaled version of our particle system given as
\begin{align}\label{eq: rescaled cont.}
    \bm{Z}_t^N = (\bm{\theta}^N_t, N^{-1/2} \bm{X}_t^{1, N}, \ldots, N^{-1/2} \bm{X}_t^{N,N}).
\end{align}
which proves much easier to analyse, and causes no loss in precision since we shall be considering only the convergence of the $\theta_n$ component. 

Our main result can be summarised as follows: recall that our main aim is to show that the sequence $(\theta_n)_{n\geq 1}$ \textit{optimises} $k(\theta)$, in other words $\theta_n$ converges to $\bar{\theta}^\star$ (which is unique under our strong convexity assumptions; see Section~\ref{sec: analysis}). Using the triangle inequality we obtain the following \textit{splitting} of the optimisation error into three terms
\begin{align*}
\mathbb{E}\left[\Vert \theta_n-\bar{\theta}^\star \Vert^2\right]^{1/2} &= W_2(\delta_{\bar{\theta}^\star}, \mathcal{L}(\theta_n))\\
&\leq W_2(\delta_{\bar{\theta}^\star}, \pi^N_\Theta)+ W_2(\pi^N_\Theta, \mathcal{L}(\bm{\theta}^N_{n\gamma}))+W_2(\mathcal{L}(\bm{\theta}^N_{n\gamma}), \mathcal{L}_{\theta_0}(\theta_n)),
\end{align*}
where each term is separately bounded in terms of the key parameters of the problem. In particular, the first term is handled via a concentration result about the $\theta$-marginal of the target, the second term is the exponential convergence of the diffusion to the target, and finally the last term is the discretisation error.

More explicitly, we prove in Theorem~\ref{thm:main} that, under appropriate regularity conditions, $\theta$-iterates in Algorithm~\ref{algo:ipsem} satisfy the following nonasymptotic convergence result
\begin{equation*}
\mathbb{E}\left[\Vert \theta_n-\bar{\theta}^\star \Vert^2 \right]^{1/2}\leq \sqrt{\frac{2 d_\theta}{N\mu}}+e^{-\mu n \gamma} \left(C_0 + \left(\frac{d_xN + d_\theta}{N \mu}\right)^{1/2}\right) + C_1 (1+\sqrt{d_\theta/N + d_x})\gamma^{1/2},
\end{equation*}
where $\mu$ is related to the convexity properties that we assume (see Section~\ref{sec: analysis}), $\gamma$ is the time discretisation parameter, $N$ is the number of particles,
and $C_1$ is a constant independent of $N,n, d_\theta, d_x, \gamma$ defined for $\gamma$ sufficiently small. Furthermore, under the additional smoothness assumption \Cref{Smoothness assump} one has the following error with improved order of numerical convergence
\begin{equation*}
\mathbb{E}\left[\Vert \theta_n-\bar{\theta}^\star \Vert^2 \right]^{1/2}\leq \sqrt{\frac{2 d_\theta}{N\mu}}+e^{-\mu n \gamma} \left(C_0 + \left(\frac{d_xN + d_\theta}{N \mu}\right)^{1/2}\right) + C_2 (1+d_\theta/N + d_x)\gamma,
\end{equation*}
This result is the first of its kind for the IPS based methods for the MMLE problem. We now turn to the proofs of our results.

\section{Analysis}\label{sec: analysis}
We first provide essential assumptions that will be central for our main results. We will also introduce further smoothness assumption to refine these results a bit later.

First, we assume that $\nabla U(\theta,x)$ is Lipschitz in both variables.
\begin{assumptionA}\label{Lipschitz assump} Let $v = (\theta, x)$ and $v' = (\theta', x')$. We assume that $U \in C^1(\mathbb{R}^{d_\theta+d_x})$ and that there exists $L>0$ such that
\begin{align*}
    \|\nabla U(v) - \nabla U(v')\| \leq L\|v-v'\| .
\end{align*}
\end{assumptionA}
We also assume that the following strong convexity condition holds.
\begin{assumptionA} \label{Conv assump} Let $v = (\theta, x)$. Then, there exists $\mu>0$ such that
\begin{align*}
 \left\langle v - v', \nabla U(v) - \nabla U(v')\right\rangle \geq     \mu \|v - v'\|^2,
\end{align*}
for all $v, v^\prime\in\mathbb{R}^{d_\theta}\times\mathbb{R}^{d_x}$.
\end{assumptionA}

\begin{assumptionA} \label{Moments assump} There is a constant $H>0$ such that the initial condition $Z^N_0=(\theta_0, N^{-1/2}X^{1,N}_0, ..., N^{-1/2}X^{N,N}_0)$ satisfies $\mathbb{E} [\Vert Z^N_0 \Vert^2] \leq H$.
\end{assumptionA}
\Cref{Lipschitz assump} is common in the literature on interacting particle systems and guarantees that both the interacting particle system \eqref{eq:ContIPS_theta}--\eqref{eq:ContIPS_x} and its time discretisation \eqref{eq:IPS_disc_theta}--\eqref{eq:IPS_disc_x} are stable (see, e.g., \cite{sznitman1991topics, bossy1997stochastic}).
\Cref{Conv assump} guarantees that $U$ has a unique minimiser $v^\star = (\theta^\star, x^\star)$, and  \Cref{Moments assump} ensures that the continuous time and discretised processes both have finite second moment for all times.
\begin{rmrk}
We note that the strong monotonicity assumption \Cref{Conv assump} on $\nabla U$ is equivalent to the assertion that $U$ is strongly convex, which in turns implies a quadratic lower bound on the growth of $U$. Therefore $e^{-U}$ is absolutely integrable, and so the Leibniz integral rule for differentiation under the
integral sign (e.g. \cite[Theorem 16.8]{billingsley1995measure}) guarantees that $k$ given in \eqref{eq:def:k} satisfies
\begin{align}\label{eq: derivative of k}
    \nabla k(\theta) = \nabla \int_{\mathbb{R}^{d_x}} e^{-U(\theta,x)}d x= -\int_{\mathbb{R}^{d_x}} \nabla_\theta U(\theta, x)e^{-U(\theta,x)}d x.
\end{align}
\end{rmrk}

\begin{example}
Consider the toy LVM of \cite{kuntz2023particle}, given by $x|\theta  \sim  \mathcal{N}(\cdot;\theta\textsf{1}_{d_x}, \textsf{Id}_{d_x})$ and $ y|x \sim \mathcal{N}(\cdot;x, \textsf{Id}_{d_x})$, where $\textsf{Id}_{d}$ denotes the identity matrix of size $d$. In this case $$\nabla U(\theta, x) = \left(d_x\theta-\sum_{i=1}^{d_x}x_i, 2x_1-\theta-y_1, \dots, 2x_{d_x}-\theta-y_{d_x}\right).$$
Since $\nabla U(\theta, x)$ is linear both in $\theta$ and in $x$ it follows that it Lipschitz continuous with constant $L=\max\{d_x+1, 3\}$; it is also strongly convex since the smallest eigenvalue of $\nabla^2 U(\theta, x)$ is given by $(2+d_x-\sqrt{4+d_x^2})/2\geq 1/4$ when $d_x\geq 1$.
\end{example}
\subsection{Continuous-time process and its invariant measure}
Recall our continuous-time interacting SDE introduced in Eqs.~\eqref{eq:ContIPS_theta}--\eqref{eq:ContIPS_x}
\begin{equation}
\label{eq: IPS noised}
    d \bm{\theta}^N_t = -\frac{1}{N}\sum_{j=1}^N \nabla_{\theta} U(\bm{\theta}^N_t, \bm{X}_t^{j, N})d t+ \sqrt {\frac{2}{N}}d \bm{B}_t^{0,N}, \qquad d \bm{X}_t^{i, N} = -\nabla_x U(\bm{\theta}^N_t, \bm{X}_t^{i, N})d t + \sqrt{2}d \bm{B}_t^{i, N},
\end{equation}
In what follows, we show that~\eqref{eq: IPS noised} admits a unique solution under~\Cref{Lipschitz assump}.
\begin{prpstn}
\label{prop:e!} Let \Cref{Lipschitz assump} and \Cref{Moments assump} hold. Then there exists a unique strong solution to~\eqref{eq: IPS noised}.
\end{prpstn}
\begin{proof}
See Appendix \ref{app:ergodicity}.
\end{proof}
Having shown that~\eqref{eq: IPS noised} admits a unique solution, we now investigate its invariant measure.
\begin{prpstn}[Invariant measure]
\label{prop:invariant_measure}
For any $N\in\mathbb{N}$, the measure $\pi_\star^N(\theta, x_1, ..., x_N) \propto e^{-\sum^N_{i=1} U(\theta, x_i)}$ is an invariant measure for the IPS~\eqref{eq: IPS noised}.
\end{prpstn}
\begin{proof}
See Appendix \ref{app:ergodicity}.
\end{proof}
\subsection{Concentration around the minimiser}
Proposition~\ref{prop:invariant_measure} shows that the IPS~\eqref{eq: IPS noised} has an invariant measure which admits
\begin{equation}
\label{eq:mu_marginal_theta}
\pi^N_\Theta (\theta) \propto \int_{\mathbb{R}^{d_x}}... \int_{\mathbb{R}^{d_x}} e^{-\sum^N_{i=1} U(\theta, x_i)} d x_1 d x_2 \dots d x_N = \biggr(\int_{\mathbb{R}^{d_x}} e^{-U(\theta, x)} d x \biggr)^N = k(\theta)^N
\end{equation}
as $\theta$-marginal. This observation is key to showing that the IPS \eqref{eq: IPS noised} can act as a \textit{global optimiser} of $k(\theta)$, or more precisely $\log k(\theta)$. Let $K(\theta) = -\log k(\theta)$ and note
\begin{align}\label{eq:target_kappa}
    \pi_\Theta^N(\theta) \propto e^{-N K(\theta)},
\end{align}
which concentrates around the minimiser of $K(\theta)$, hence the maximiser of $k(\theta)$, as $N \to \infty$. This is a classical setting in global optimisation, see, e.g., \cite{hwang1980laplace} where $N$ acts as the \textit{inverse temperature} parameter. Here we stress that our number of particles produces an equivalent effect as the classical inverse temperature parameter, hence we do not need an explicit inverse temperature parameter in this setting. In the next proposition we provide quantitative rates on how $\pi^N_{\Theta}$ concentrates around the minimiser of~\eqref{eq:def:k}. Note that concentration results of this kind hold in more general contexts, see Proposition 3.4 of \cite{pmlr-v65-raginsky17a} or see \cite{zhang2023nonasymptotic}.

\begin{prpstn}[Concentration Bound]
\label{prop:concentration}
Let $\pi^N_\Theta$ be as in \eqref{eq:mu_marginal_theta} and $\bar{\theta}^\star$ be the maximiser of $k(\theta)$. Then, under~\Cref{Conv assump}, one has the bound
\begin{equation*}
W_2(\pi^N_\Theta, \delta_{\bar{\theta}^\star})\leq \sqrt{\frac{2d_\theta}{\mu N}}.
\end{equation*}
\end{prpstn}
\begin{proof}
See Appendix \ref{app:ergodicity}.
\end{proof}

\subsection{Convergence of theta-marginal}
We now exploit the connection between~\eqref{eq: IPS noised} and standard Langevin diffusions to establish exponential ergodicity of~\eqref{eq: IPS noised}.

As mentioned before, recall that we can define the rescaled version of our particle system
\begin{align}\label{eq: rescaled Z 2}
    \bm{Z}_t^N = (\bm{\theta}^N_t, N^{-1/2} \bm{X}_t^{1, N}, \ldots, N^{-1/2} \bm{X}_t^{N,N})
\end{align}
in which the $\theta$-coordinate is not modified while the $X$-coordinate is rescaled by $N^{-1/2}$.
With the help of this rescaled IPS, we have the following exponential ergodicity result:
\begin{prpstn} \label{prop: conv to inv} Let \Cref{Lipschitz assump}, \Cref{Conv assump} and \Cref{Moments assump} hold. Then, for any $N\in\mathbb{N}$ one has
\begin{align*}
    W_2(\mathcal{L}(\bm{\theta}^N_t), \pi_{\Theta}^N) \leq e^{-\mu t} \left( \mathbb{E} [\|Z^N_0 - z^\star\|^2]^{1/2} + \left(\frac{d_xN + d_\theta}{N \mu}\right)^{1/2}\right),
\end{align*}
where $\mu$ is given in~\Cref{Conv assump}, $Z^N_0$ is given in \Cref{Moments assump}, $v^\star=(\theta^\star, x^\star)$ is the minimiser of $U$, and
\begin{align*}
    z^\star &= (\theta^\star, N^{-1/2} x^\star, \ldots, N^{-1/2} x^\star).
\end{align*}
\end{prpstn}
\begin{proof}
See Appendix \ref{app:ergodicity}.
\end{proof}
We note that the proof uses the contraction of the rescaled particle system, which, as we show in the proof, implies the convergence of the $\theta$-marginal (since the $\theta$-coordinate is not modified).

\subsection{Discretisation error}\label{subsec: disc error}
The final part of analysis will consist of discretisation analysis of the discrete-time scheme \eqref{eq:IPS_disc_theta}--\eqref{eq:IPS_disc_x}, i.e., the error analysis of this scheme in comparison to the continuous time process \eqref{eq: IPS noised}.
Given $\nabla U$ is Lipschitz, the $L^2$ convergence rates are of the expected order in $\gamma$, however we present full arguments in order to demonstrate that our bounds do not get worse with the number of particles $N$ (and in fact the dependence on $d_\theta$ improves as $N\to\infty$).

For the sake of proving convergence bounds for the discretisation it is convenient to extend the original definition of the discretisation $(\theta_n, X^{1,N}_n,..., X^{N,N}_n)$ to a time continuous process $(\bar{\theta}_t, \bar{X}^{1,N}_t,..., \bar{X}^{N,N}_t)$ that satisfies $\bar{\theta}_{\gamma n}=\theta_n$. Specifically, we define
\begin{align}\label{eq: interpolation}
\bar{\theta}_{t} &= \theta_0-\frac{1}{N}\sum_{j=1}^N\int^{t}_0  \nabla_{\theta} U(\bar{\theta}_{\kappa_\gamma(s)}\bar{X}_{\kappa_\gamma(s)}^{j, N}) d s+\sqrt{\frac{2}{N}}\int^{t}_0 d\bm{B}^{0,N}_s, \\
\bar{X}_t^{i, N} &= X_0^{i, N} - \int^{t}_0 \nabla_x U(\bar{\theta}_{\kappa_\gamma(s)}, \bar{X}_{\kappa_\gamma(s)}^{i, N}) d s + \sqrt{2}\int^{t}_0 d\bm{B}_s^{i, N}.
\end{align}
where $\kappa_\gamma$ is the backwards projection onto the grid $\{0, \gamma, 2\gamma, ...\}$, that is, $\kappa_\gamma(t)=\gamma\lfloor \gamma^{-1} t \rfloor$ and in particular $\kappa_\gamma(n\gamma +\varepsilon)=n$ for $n \in \mathbb{N}$ and $\varepsilon \in (0,\gamma)$. Noting then that $\bar{\theta}_t$ is an estimate for $\bm{\theta}^N_{ t}$, we can bound the error between this continuous time process and~\eqref{eq: IPS noised}. 

\begin{prpstn} \label{prop: discr error 1}
Let \Cref{Lipschitz assump},~\Cref{Conv assump}, \Cref{Moments assump} hold. Then for every $\gamma_0 \in (0, \min\{L^{-1}, 2\mu^{-1}\})$ there exists a constant $C>0$ independent of $t,n,N,\gamma, d_\theta, d_x$ such that for every $\gamma \in (0, \gamma_0)$ one has
\begin{equation*} 
\mathbb{E}\left[\Vert \theta_n -\bm{\theta}^N_{n\gamma} \Vert^2\right]^{1/2} \leq C (1+\sqrt{d_\theta/N+d_x})\gamma^{1/2},
\end{equation*}
for all $n\in\mathbb{N}$.
\end{prpstn}
\begin{proof}
See Appendix \ref{app:ergodicity}.
\end{proof}
Finally we can introduce a smoothness assumption that shall allow us to obtain a better convergence rate for the discretisation error.
\begin{assumptionA} \label{Smoothness assump} Let $U \in C^2(\mathbb{R}^{d_\theta + d_x})$ and there exists $l>0$ such that
\begin{align*}
\Vert \nabla ^2U(v)-\nabla ^2U(v') \Vert \leq l \Vert v-v' \Vert,
\end{align*}
for all $v, v^\prime\in\mathbb{R}^{d_\theta}\times\mathbb{R}^{d_x}$.
\end{assumptionA}
Under the smoothness assumption \Cref{Smoothness assump} one can improve the numerical error to $O(\gamma)$, at the cost of worsened dimensional dependence.
\begin{prpstn} \label{prop: discr error 2}
Let \Cref{Lipschitz assump},~\Cref{Conv assump}, \Cref{Moments assump} and \Cref{Smoothness assump} hold. Then for every $\gamma_0 \in (0, \min\{L^{-1}, 2\mu^{-1}, 1\})$ there exists a constant $C>0$ independent of $t,n,N,\gamma, d_\theta, d_x$ such that for every $\gamma \in (0, \gamma_0)$ one has
\begin{equation} \label{eq: Lipschtiz disc error}
\mathbb{E}\left[\Vert \theta_n -\bm{\theta}^N_{n\gamma} \Vert^2\right]^{1/2} \leq C (1+d_\theta/N+d_x)\gamma,
\end{equation}
for all $n\in\mathbb{N}$.
\end{prpstn}
\begin{proof}
See Appendix \ref{app:ergodicity}.
\end{proof}

\subsection{Global error}

Combining the results obtained in the previous sections, we are able to give precise bounds on the accuracy of Algorithm~\ref{algo:ipsem} in terms of $N$, $\gamma$, $n$ and the convexity properties of $U$.
\begin{thrm}
\label{thm:main}
Let \Cref{Lipschitz assump},~\Cref{Conv assump}, \Cref{Moments assump} hold. Then for every $\gamma_0 \in (0, \min\{L^{-1}, 2\mu^{-1}\})$ there exists a constant $C_1 > 0$ independent of $t,n,N,\gamma, d_\theta, d_x$ such that for every $\gamma \in (0, \gamma_0)$ one has
\begin{align*}
\mathbb{E}\left[\Vert \theta_n-\bar{\theta}^\star \Vert^2\right]^{1/2} \leq \sqrt{\frac{2 d_\theta}{N\mu}}+e^{-\mu n \gamma} \left(\mathbb{E} [\|Z^N_0 - z^\star\|^2]^{1/2} + \left(\frac{d_xN + d_\theta}{N \mu}\right)^{1/2}\right)+C_1 (1+\sqrt{d_\theta/N +d_x})\gamma^{1/2},
\end{align*}
for all $n\in\mathbb{N}$. Furthermore, if one additionally assumes \Cref{Smoothness assump} and $\gamma \leq 1$ then one has the bound
\begin{align*}
\mathbb{E}\left[\Vert \theta_n-\bar{\theta}^\star \Vert^2\right]^{1/2} \leq \sqrt{\frac{2 d_\theta}{N\mu}}+e^{-\mu n \gamma} \left( \mathbb{E} [\|Z^N_0 - z^\star\|^2]^{1/2} + \left(\frac{d_xN + d_\theta}{N \mu}\right)^{1/2}\right)+C_2 (1+d_\theta/N +d_x)\gamma,
\end{align*}
for $C_2 > 0$ another constant independent of $t,n,N,\gamma, d_\theta, d_x$.
\end{thrm}
\begin{proof}
Let us denote by $\mathcal{L}(\theta_n)$ the law of $\theta_n$ obtained with Algorithm~\ref{algo:ipsem}. Then, we can decompose the expectation into a term describing the concentration of the $\pi^N_\Theta$ around $\bar{\theta}^\star$, a term describing the convergence of~\eqref{eq: IPS noised} to its invariant measure, and a term describing the error induced by the time discretisation:
\begin{align*}
\mathbb{E}\left[\Vert \theta_n-\bar{\theta}^\star \Vert^2\right]^{1/2} &= W_2(\delta_{\bar{\theta}^\star}, \mathcal{L}_{\theta_0}(\theta_n))\\
&\leq W_2(\delta_{\bar{\theta}^\star}, \pi^N_\Theta)+ W_2(\pi^N_\Theta, \mathcal{L}(\bm{\theta}^N_{n\gamma}))+W_2(\mathcal{L}(\bm{\theta}^N_{n\gamma}), \mathcal{L}_{\theta_0}(\theta_n))\\
&\leq \sqrt{\frac{2 d_\theta}{N\mu}}+e^{-\mu n \gamma} \left( \mathbb{E} [\|Z^N_0 - z^\star\|^2]^{1/2} + \left(\frac{d_xN + d_\theta}{N \mu}\right)^{1/2}\right) + C_1(1+\sqrt{d_\theta/N +d_x})\gamma^{1/2},
\end{align*}
where the first equality stems from the fact that $\delta_{\bar{\theta}^\star}$ is the law of a constant random variable, the second inequality is a triangle inequality for Wasserstein distances, and the last inequality follows combining Proposition~\ref{prop:concentration}, Proposition~\ref{prop: conv to inv} and Proposition~\ref{prop: discr error 1}. For the second bound one simply replaces the bound for $W_2(\mathcal{L}(\bm{\theta}^N_{n\gamma}), \mathcal{L}_{\theta_0}(\theta_n))$ given by Proposition \ref{prop: discr error 1} by the bound given by Proposition \ref{prop: discr error 2}.
\end{proof}

\subsection{Error Bounds}
\label{sec:bounds}
Observe that Theorem~\ref{thm:main} permits the following bounds for $\mathbb{E}\left[\Vert \theta_n-\bar{\theta}^\star \Vert^2\right]^{1/2}=\mathcal{O}(\varepsilon)$ in terms of the key parameters $d_\theta, d_x$-
\begin{center}
\begin{tabular}{ |c ||c |c |c|}
\hline
 $\mathcal{O}(\varepsilon)$ & $N$ & $\gamma$ & $n$  \\ 
 \hline \hline
 \Cref{Lipschitz assump}, \Cref{Conv assump}, \Cref{Moments assump}  & $\mathcal{O}(d_\theta \varepsilon^{-2})$ & $\mathcal{O}(d_x^{-1}\varepsilon^2)$ & $\mathcal{O}(d_x \varepsilon^{-2-\delta})$\\
 \hline
 \Cref{Lipschitz assump}, \Cref{Conv assump}, \Cref{Moments assump}, \Cref{Smoothness assump} & $\mathcal{O}(d_\theta \varepsilon^{-2})$ &  $\mathcal{O}(d_x ^{-1}\varepsilon)$& $\mathcal{O}(d_x \varepsilon^{-1-\delta})$ \\  
 \hline
\end{tabular}
\end{center}
where $\delta>0$ is any small positive constant. These bounds follow from first choosing $N$ so that the first term is $\mathcal{O}(\epsilon)$ and $\gamma$ sufficiently small to counteract the dependence on $d_x$ in the third term. Finally, since for every $p\in \mathbb{N}$ one has $e^x\geq \frac{1}{p!}x^p$ for $x>0$, for every $\delta>0$ (by choosing $p \in \mathbb{N}$ large enough) one has $e^{-\epsilon^\delta}\leq C\epsilon$. Therefore, as long as $n$ is chosen sufficiently large that $\mu n \gamma = \mathcal{O}(\varepsilon^{-\delta})$, the exponential decay is strong enough that the middle term is of order $\mathcal{O}(\varepsilon)$.

For every step of the algorithm one requires $2N$ evaluations of $\nabla U$,  $2N(d_\theta+d_x)$ evaluations of $\nabla U$ component-wise, and $d_\theta+Nd_x$ independent standard $1$-dimensional Gaussians. Therefore, one has the following table with the number of evaluations of $\nabla U$ and number of independent $1$d Gaussians that need to be called
\begin{center}
\begin{tabular}{ |c ||c |c |c|}
\hline
$\mathcal{O}(\varepsilon)$ &  Evaluations of $\nabla U$ &  Evaluations of $(\nabla U)^j$ & Independent $1$d Gaussians \\ 
 \hline \hline
 \Cref{Lipschitz assump}, \Cref{Conv assump}, \Cref{Moments assump}  & $\mathcal{O}(d_\theta d_x \varepsilon^{-4-\delta})$ & $\mathcal{O}(d_\theta d_x(d_\theta+d_x)\varepsilon^{-4-\delta})$ &  $\mathcal{O}(d_\theta d_x^2\varepsilon^{-4-\delta})$\\
 \hline
 \Cref{Lipschitz assump}, \Cref{Conv assump}, \Cref{Moments assump},  \Cref{Smoothness assump} & $\mathcal{O}(d_\theta d_x \varepsilon^{-3-\delta})$  & $\mathcal{O}(d_\theta d_x(d_\theta+d_x)\varepsilon^{-3-\delta})$ &  $\mathcal{O}(d_\theta d_x^2\varepsilon^{-3-\delta})$\\ 
 \hline
\end{tabular}
\end{center}

\section{Extension to stochastic gradients}\label{sec: stochastic gradient}
We also extend our analysis to stochastic gradients. This is especially useful in the context of machine learning, where the gradient of the loss function is often estimated using a mini-batch of data \cite{bottou2018optimization}. It is also useful in the context of Bayesian inference, where the gradient of the log-posterior is often estimated using a mini-batch of data \cite{welling2011bayesian} as well as in variational inference \cite{hoffman2013stochastic}.

To unlock the applicability of our results in these settings, we analyse Algorithm \ref{algo:ipsem} when the gradients $\nabla_\theta U$, $\nabla _x U$ are replaced with unbiased estimators with finite variance. Specifically, we define the process $(\tilde{\theta}_n, \tilde{X}^{1,N}_n,...,\tilde{X}^{N,N}_n)$ by
\begin{align}
    \label{eq: theta stoch} \tilde{\theta}_{n+1} &= \tilde{\theta}_n - \frac{\gamma}{N}\sum_{j=1}^N h_\theta(\tilde{\theta}_n, \tilde{X}_n^{j, N}, q_{n+1}) + \sqrt {\frac{2\gamma}{N}}\xi_{n+1}^{0,N}, \;\;\;  \tilde{\theta}_0=\theta_0  \\
    \label{eq: X stoch} \tilde{X}_{n+1}^{i, N} &= X_n^{i, N} - \gamma h_x (\tilde{\theta}_n, \tilde{X}_n^{i, N}, q_{n+1}) + \sqrt{2\gamma} \xi_{n+1}^{i, N}, \;\;\; \tilde{X}^{i,N}_0=X^{i,N}_0, \;\;\; i=1,2,...,N
\end{align}
where $(q_n)_{n=1}^\infty$ is a sequence of random variables and $h=(h_x,h_\theta):\mathbb{R}^{d_\theta}\times\mathbb{R}^{d_x}\times \mathbb{R}\to\mathbb{R}^{d_\theta}\times\mathbb{R}^{d_x}$ is a function satisfying the following assumption.
\begin{assumptionB}\label{Stoch Gradient assump}
The sequence of independent $\mathbb{R}$-valued random variables $(q_n)_{n=1}^\infty$ is independent of the noise $(\bm{B}_t)_{t\geq0}$ and the initial condition $(\theta_0, X^{1,N}_0,..., X^{N,N}_0)$. Furthermore, there exists a constant $m>0$ such that for every $v\in\mathbb{R}^{d_\theta}\times\mathbb{R}^{d_x}$ and $n\in\mathbb{N}$
\begin{align*}
    \mathbb{E} [h(v,q_n)]=\nabla U(v)\qquad\qquad
    \mathbb{E} [\Vert h(v,q_n)\Vert^2] \leq m(1+\Vert v \Vert^2).
\end{align*}

\end{assumptionB}
\vspace{5mm}
\begin{rmrk}\label{remark: restriction on m}
    Note that by the monotonicity assumption, for $v \in \mathbb{R}^{d_\theta+d_x}$ and $v^\star = (\theta^\star, x^\star)$ one has
  \begin{equation}
 \mathbb{E}[\langle  h(v,q_n), v-v^\star\rangle] =\mathbb{E}[\langle  h(v,q_n)-h(v^\star,q_n), v-v^\star \rangle] \geq \mu \Vert v-v^\star\Vert^2,
  \end{equation}
  so that one must have $\mathbb{E}[\Vert h(v,q_n)\Vert] \geq \mu \Vert v-v^\star\Vert$, and therefore $m\geq \mu^2$.
\end{rmrk}
This then allows us to achieve a result comparable to Theorem \ref{thm:main} in the presence of gradient noise.
\begin{thrm}
\label{thm:main stoch}
%
Let \Cref{Lipschitz assump},~\Cref{Conv assump}, \Cref{Moments assump}, \Cref{Stoch Gradient assump} hold. Then for every $\gamma_0\in(0,\frac{\mu}{2m})$ there exists a constant $C_3 > 0$ independent of $t,n,N,\gamma, d_\theta, d_x$ such that for every $\gamma \in (0, \gamma_0)$ one has
\begin{align*}
\mathbb{E}\left[\Vert \tilde{\theta}_n-\bar{\theta}^\star \Vert^2\right]^{1/2} \leq \sqrt{\frac{2 d_\theta}{N\mu}}+e^{-\mu n \gamma} \left(\mathbb{E} [\|Z^N_0 - z^\star\|^2]^{1/2} + \left(\frac{d_xN + d_\theta}{N \mu}\right)^{1/2}\right)+C_3 (1+\sqrt{d_\theta/N +d_x})\gamma^{1/2},
\end{align*}
for all $n\in\mathbb{N}$. 
\end{thrm}
\begin{proof}
The proof is the same as in Theorem \ref{thm:main}, but replacing Proposition \ref{prop: discr error 1} with Proposition \ref{prop: stoch discr error}.
\end{proof}

\section{Maximum marginal likelihood estimation with gradient flows}
\label{sec:related}
Our approach is based on observing that in order to maximise the marginal likelihood~\eqref{eq:def:k} one can employ standard tools from Langevin-based optimisation. \cite{neal1998view} and, more recently, \cite{kuntz2023particle} observe that EM can be seen as application of coordinate ascent to the free-energy functional $F:\mathbb{R}^{d_\theta}\times\mathcal{P}(\mathbb{R}^{d_x})\to \mathbb{R}$
\begin{align*}
    F(\theta, \nu) := - \int U(\theta, x)\nu( x)d x - \int \log\left( \nu(x)\right)\nu( x)d x.
\end{align*}
The approach of \cite{kuntz2023particle} leverages recent advances in minimisation of functionals over spaces of probability measures \cite{otto2001geometry, jordan1998variational} to obtain an IPS approximating the minimiser $(\bar{\theta}^\star, \nu^\star)\in \mathbb{R}^{d_\theta}\times\mathcal{P}(\mathbb{R}^{d_x})$ of $F$. In this setting, $\bar{\theta}^\star$ corresponds to the MMLE estimate, while $\nu^\star = p_{\bar{\theta}^\star}(x|y)$ is the posterior
distribution of the latent variable $x$ given the observed data $y$.
\cite[Theorem 1--2]{kuntz2023particle} establish that minimising $k$ in~\eqref{eq:def:k} is equivalent to minimising $F$.

The minimisation of $F$ is achieved by deriving a Wasserstein gradient flow for $F$, which can be seen as an extension of standard gradient flows to spaces of probability measures.
The Wasserstein gradient flow for $F$ over $\mathbb{R}^{d_\theta}\times\mathcal{P}(\mathbb{R}^{d_x})$ is given by the following combination of a gradient descent ODE for $\theta$ and a Fokker-Planck PDE for $\nu$
\begin{align*}
    d \bm{\theta}_t &= \left[-\int_{\mathbb{R}^{d_x}} \nabla_{\theta} U(\bm{\theta}_t, x)\nu_t(x)d x\right] d t\\
    \partial_t\nu_t &= \nabla_x\cdot\left(\nu_t\nabla_x\log\frac{p_\theta(\cdot, y)}{\nu_t}\right).
\end{align*}


The associated SDE is a McKean--Vlasov SDE (MKVSDE), whose drift coefficient depends on the law $\nu_t$ of the solution $\bm{X}_t$
\begin{align} \label{eq: Cont EM}
    d \bm{\theta}_t = \left[-\int_{\mathbb{R}^{d_x}} \nabla_{\theta} U(\bm{\theta}_t, x)\nu_t(x)d x\right] d t, \qquad d \bm{X}_t = -\nabla_x U(\bm{\theta}_t, \bm{X}_t)d t + \sqrt{2}d \bm{B}_t,
\end{align}
where $(\bm{B}_t)_{t \geq 0}$ is a $d_x$-dimensional Brownian motion.
As shown in \cite[Theorem 3]{kuntz2023particle}, under the strong convexity assumption \Cref{Conv assump}, \eqref{eq: Cont EM} will converge
exponentially fast to $\bar{\theta}^\star$ and the corresponding posterior distribution $p_{\bar{\theta}^\star}(x|y)$.

The SDE~\eqref{eq: Cont EM} gives rise to an IPS closely related to that in~\eqref{eq: IPS noised}
\begin{equation}
\label{eq:juan_sde}
    d \bm{\theta}^N_t = -\frac{1}{N}\sum_{j=1}^N \nabla_{\theta} U(\bm{\theta}^N_t, \bm{X}_t^{j, N})d t, \qquad d \bm{X}_t^{i, N} = -\nabla_x U(\bm{\theta}^N_t, \bm{X}_t^{i, N})d t + \sqrt{2}d \bm{B}_t^{i, N},
\end{equation}
for $i=1, \dots, N$, where $\{(\bm{B}_t^{i,N})_{t \geq 0}\}_{i =0}^N$ is a family of independent
Brownian motions. 
The main difference between~\eqref{eq: IPS noised} and the MKVSDE above is that the $\theta$-component evolves as an ODE rather than an SDE, as a consequence the system~\eqref{eq:juan_sde} does not have $\pi^N_\star$ as invariant measure, nor it can be easily cast as a Langevin diffusion targeting a given distribution.

Despite this difference, the IPS~\eqref{eq: IPS noised} and~\eqref{eq:juan_sde} are closely related. Under~\Cref{Conv assump}, standard results on the $N\to \infty$ limit of interacting particle systems (e.g. \cite[Theorem 3.3]{malrieu2001logarithmic}), allow us to show the following convergence of the particle system~\eqref{eq: IPS noised} to~\eqref{eq: Cont EM}.

\begin{prpstn}[Propagation of chaos]
\label{prop:poc}
Let $\bm{\theta}_t$ denote the $\theta$-component of the solution of the gradient flow ODE~\eqref{eq: Cont EM}, $\bm{\theta}^N_t$ denote the $\theta$-component of the IPS~\eqref{eq: IPS noised} and $\bm{\widetilde{X}}^{i, N}_t$ for $i=1, \dots, N$ denote $N$ independent copies of the $X$-component of~\eqref{eq: Cont EM}.
Under~\Cref{Conv assump} and assuming that $\nabla U$ has polynomial growth, for any (exchangeable) initial condition $(\theta_0^{N}, X_0^{1:N})$, we have for any $t \geq 0$
\begin{equation}
\label{eq:poc}
\sup_{t\geq 0}\mathbb{E}\left[\|\bm{\theta}^N_t - \bm{\theta}_t\|^2 +\frac{1}{N}\sum_{i=1}^N \|\bm{X}^{i, N}_t - \bm{\widetilde{X}}^{i, N}_t\|^2\right]\leq \frac{ 2d_\theta +C}{N\mu},
\end{equation}
where $C$ is a constant which depends on the Lipschitz constant of $\nabla U$.
\end{prpstn}
\begin{proof}
    See Appendix~\ref{app:gf}.
\end{proof}

A similar result can be established for~\eqref{eq:juan_sde} using standard tools from the literature on IPS (see, e.g. \cite{sznitman1991topics, bossy1997stochastic, malrieu2001logarithmic}). If follows that the addition of the noise in the $\theta$-component does not undermine the gradient flow behaviour of~\eqref{eq: IPS noised}.

\section{Experiments}\label{sec:experiments}
In this section, we provide numerical experiments to demonstrate the empirical behaviour of IPLA in relation to similar competitors, such as particle gradient descent (PGD) scheme \cite{kuntz2023particle} and the SOUL algorithm \cite{de2021efficient}.

\begin{figure}[t]
    \centering
    \includegraphics[width=\textwidth]{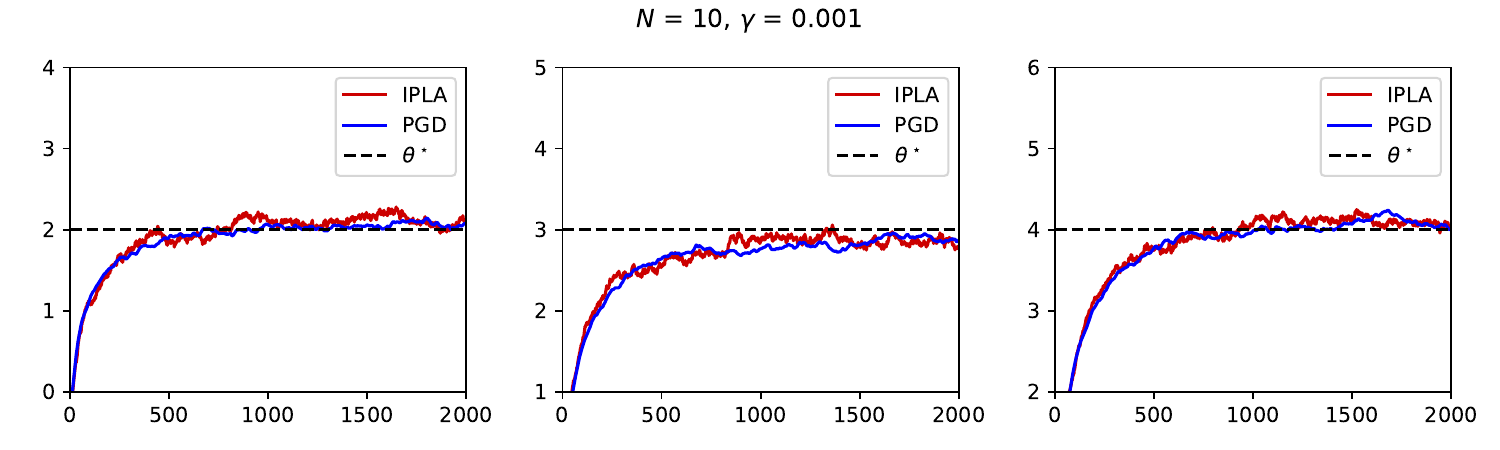}
    \includegraphics[width=\textwidth]{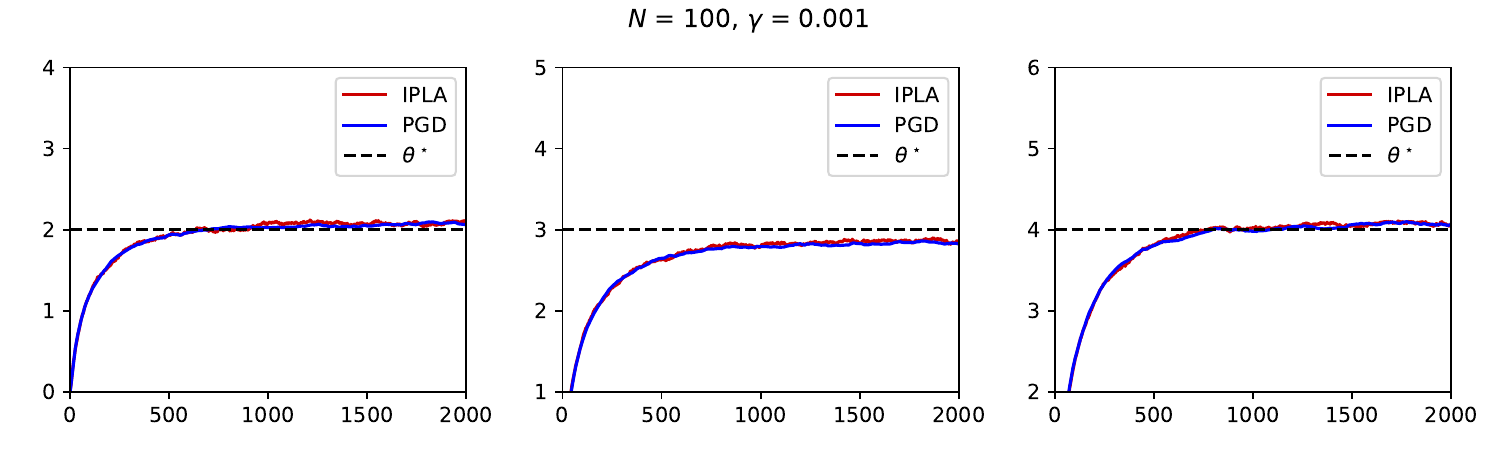}
    \includegraphics[width=\textwidth]{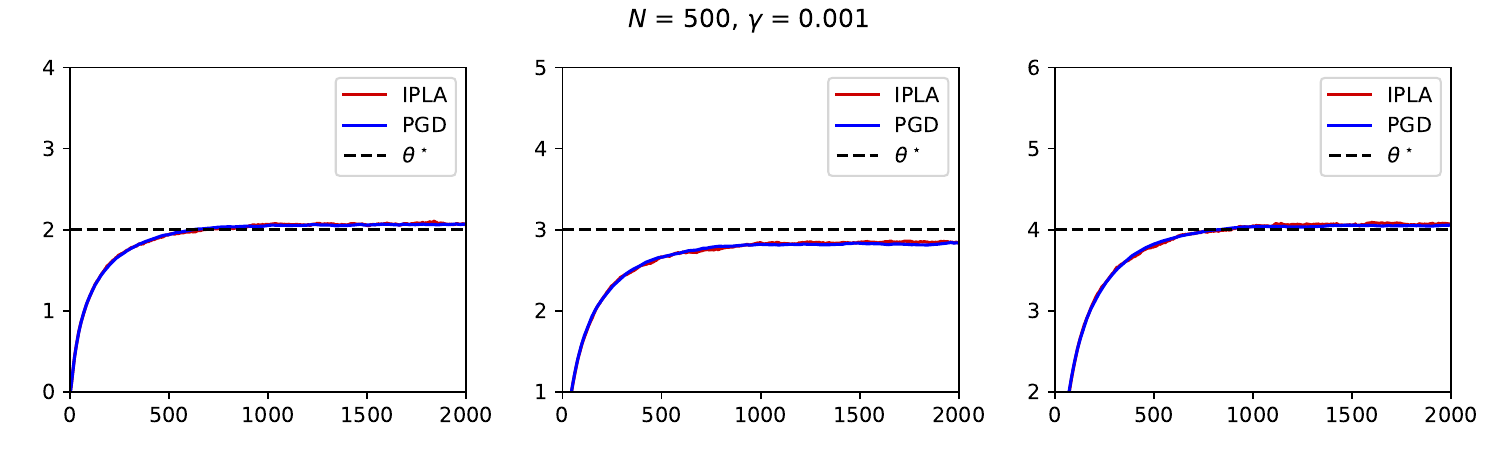}
    \caption{The performance of IPLA and PGD on the synthetic logistic regression problem. Each column corresponds to one of the components of the true parameter $\bar{\theta}^\star = [2,3,4]$. The top row corresponds to $N=10$, the middle row to $N=100$ and the bottom row to $N=500$. As expected, the PGD and IPLA perform similarly, especially as $N$ increases.}
    \label{fig:synthetic-ipla-pgd}
\end{figure}

\subsection{Bayesian logistic regression on synthetic data}
\label{sec:logistic_synth}
We first compare the method proposed by \cite{kuntz2023particle}, which is termed particle gradient descent (PGD), and IPLA on a simple Bayesian regression task. 
We consider the Bayesian logistic regression LVM where for $\theta\in\mathbb{R}^{d_\theta}$
\begin{align*}
p_\theta(x) &= \mathcal{N}(x;\theta , 
\sigma^2 \textsf{Id}_{d_x}),\qquad\qquad
p_\theta(y|x) = \prod_{j=1}^{d_y}s(v_j^Tx)^{y_j}(1-s(v_j^Tx))^{1-y_j},
\end{align*}
with $d_\theta = d_x$, $s(u):=e^u/(1+e^u)$ the logistic function and $\{v_j\}_{j=1}^{d_y}\in \mathbb{R}^{d_x}$ a set of covariates with corresponding binary responses $\{y_j\}_{j=1}^{d_y}\in \{0, 1\}$. We assume that the value of $\sigma^2$ is fixed and known.
In this section, we generate a set of synthetic $d_x$-dimensional  covariates $v_j$ for $j=1, \dots, d_y$ and then generate our synthetic data $\{y_j\}_{j=1}^{d_y}$ sampling $y_j|\theta, x, v_j$ from a Bernoulli random variable with parameter $s(v_j^Tx)$.
The marginal likelihood is given by
\begin{align*}
    k(\theta) =p_\theta(y)= (2\uppi\sigma^2)^{-d_x/2}\int_{\mathbb{R}^{d_x}} \left(\prod_{j=1}^{d_y}s(v_j^Tx)^{y_j}(1-s(v_j^Tx))^{1-y_j}\right)\exp\left(-\norm{x-\theta}^2/(2\sigma^2)\right)d x.
\end{align*}

We then identify the performance of PGD and IPLA w.r.t. to the ground truth parameter. We note that only difference between PGD and IPLA is the noise in $\theta$-dimension, hence these algorithms perform similarly on a variety of tasks. In the next section, we will aim to identify a difference in their performance on the example provided on a real dataset. Before moving to the experimental setup, we note that given the model above, one can check $U(\theta,x)$ is strictly convex and satisfies \Cref{Lipschitz assump} (see Appendix~\ref{app:verify_a1_a2} for details).

\subsubsection{Experimental Setup}
We choose $d_x = d_\theta = 3$ and $d_y = 900$ and set $\bar{\theta}^\star = [2, 3, 4]$ to test the convergence to the parameter. We run the methods for $M = 2000$ steps, with $N = 10, 100, 500$ particles and $\gamma = 0.001$. We set $\sigma = 0.01$ and simulate $v_j \sim \textsf{Unif}(-1, 1)^{\otimes d_x}$. We initialise the particles $X_0^{i, N}$ from a zero-mean Gaussian with covariance $\textsf{Id}_{d_x}$ and the initial parameter $\theta_0$ from a deterministic value $\theta_0 = [0, 0, 0]$.

We see from Fig.~\ref{fig:synthetic-ipla-pgd} that both algorithms perform the same. This is expected since the only difference between the two algorithms is the noise in the $\theta$-dimension. As $N$ increases, the performance of both algorithms improves and IPLA gets closer to PGD, since the noise term in the $\theta$-component becomes smaller than the drift term.

\subsection{Bayesian logistic regression on Wisconsin cancer data}

We next adopt a realistic example using the same setup as in \cite{kuntz2023particle_code}. In particular, in this section, we compare IPLA to PGD as well as to more standard methods like mean-field variational inference (MFVI) and SOUL \cite{de2021efficient}.

\begin{figure}
    \centering
    \includegraphics[width=0.32\textwidth]{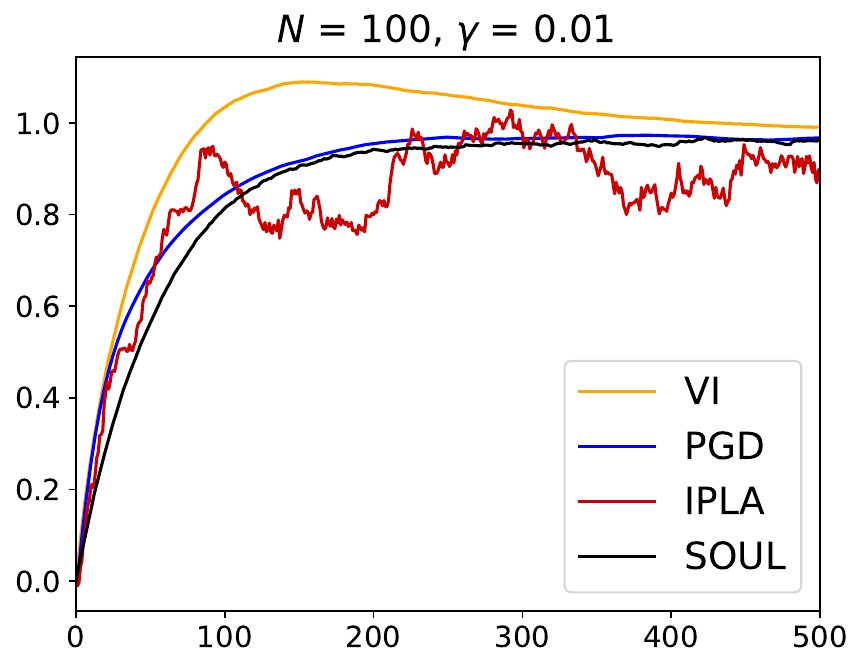}
    \includegraphics[width=0.32\textwidth]{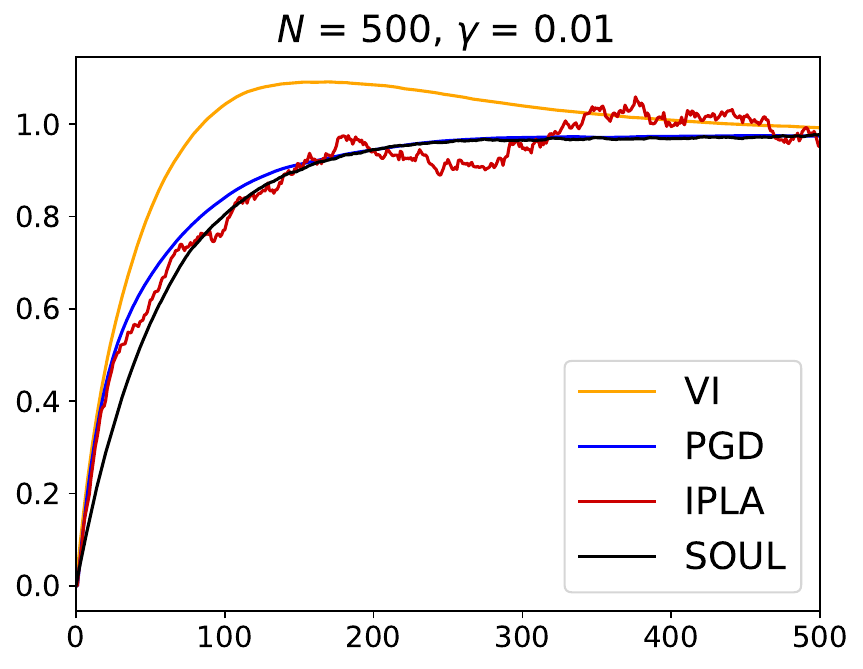}
    \includegraphics[width=0.32\textwidth]{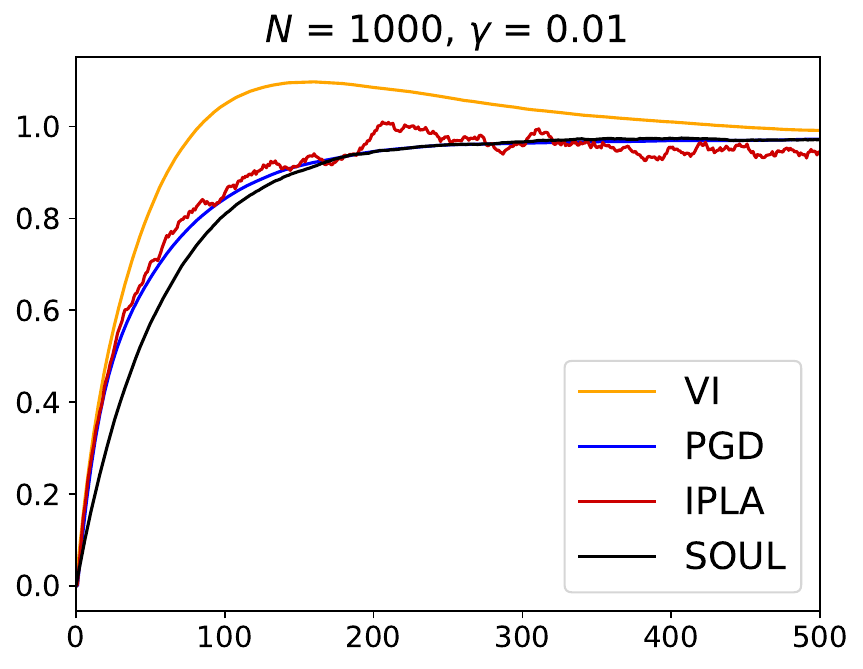}
    \caption{The performance of IPLA, PGD, MFVI and SOUL on the logistic regression experiment on real data. The left plot corresponds to $N=100$, the middle plot to $N=500$ and the right plot to $N=1000$. As expected, the PGD and IPLA perform similarly, especially as $N$ increases. We also note that SOUL is significantly slower compared to PGD and IPLA.}
    \label{fig:logistic-real}
\end{figure}

The setup in this section is a simplified version of the synthetic example in the last section, for which we verified our assumptions. Therefore, we do not verify our assumptions here again as they hold as a corollary of the proof in Sec.~\ref{sec:logistic_synth}. We use in this section a breast cancer dataset with $d_y = 683$ data points and $d_x = 9$ latent variables \cite{breast_cancer_wisconsin_diagnostic_17}. These $9$ latent variables correspond to latent ``features'' which are extracted from a digitised image of a fine needle aspirate (FNA) of a breast mass. The features describe characteristics of the cell nuclei present in the image \cite{breast_cancer_wisconsin_diagnostic_17}. The data being $y=0$ or $y=1$ corresponds to whether the mass is benign or malignant. We use the setting provided by \cite{kuntz2023particle_code} and compare our method with PGD, MFVI and SOUL. For the implementation details of SOUL, PGD, and MFVI, we refer to Appendix~E.2 of \cite{kuntz2023particle}.

For the estimation problem, we define
\begin{align*}
    p_\theta(x) &= \mathcal{N}(x;\theta \mathbf{1}_{d_x}, 5 \textsf{Id}_{d_x}),
\end{align*}
and our likelihood is
\begin{align*}
    p(y|x) &= \prod_{j=1}^{d_y}s(v_j^Tx)^{y_j}(1-s(v_j^Tx))^{1-y_j}.
\end{align*}
In this example, we do not have a ``ground truth'' value as the data is real. We set $\gamma = 0.01$ for all algorithms as this is the setup used in the fair comparison of algorithms in \cite{kuntz2023particle}. We run the algorithms for $M = 500$ steps and vary the number of particles $N \in \{100, 500, 1000\}$. We note that, under this setting, SOUL gets increasingly expensive as it requires $N$ sequential ULA steps to update the parameter once, while PGD and IPLA can parallelise this computation. \cite[Table 1]{kuntz2023particle} shows that for $N=100$ one has that SOUL is at least 10 times slower than PGD on the same example considered here; since the computational cost of IPLA and PGD are identical we do not repeat the comparison. 

\begin{figure}
    \centering
    \includegraphics[width=0.5\textwidth]{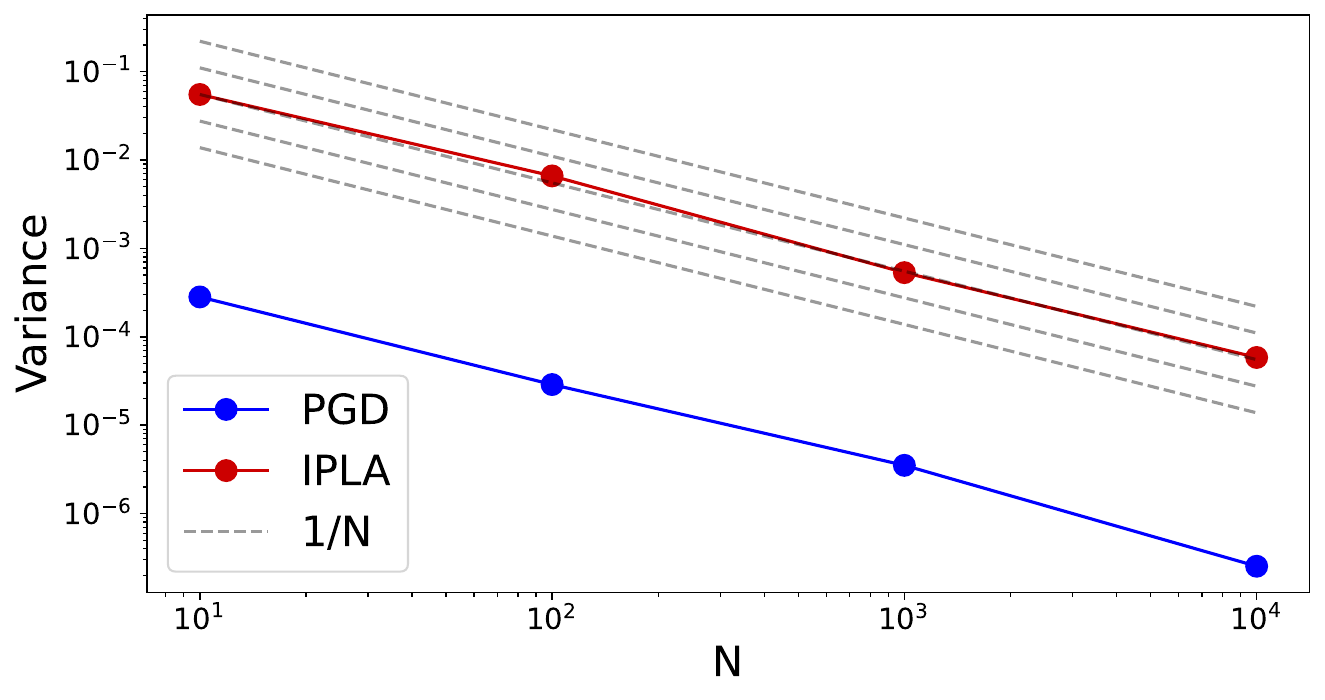}
    \caption{The convergence rate of the variance of the parameter estimates produced by PGD and IPLA over $100$ Monte Carlo runs for each $N \in \{10, 100, 1000, 10000\}$. We can see that the $\mathcal{O}(1/N)$ convergence rate holds for the second moments as suggested by our results for IPLA. 
    }
    \label{fig:logistic-real-rate}
\end{figure}

We present the results in Fig.~\ref{fig:logistic-real}, where we see that PGD and IPLA perform similarly. We also note that MFVI performs poorly compared to the other methods. This is expected as MFVI is a mean-field approximation and does not capture the correlation between the latent variables. We have also tested the variance of the estimates produced by PGD and IPLA over $100$ Monte Carlo runs for each $N \in \{10, 100, 1000, 10000\}$ and present the results in Fig.~\ref{fig:logistic-real-rate}. We see that the $\mathcal{O}(1/N)$ convergence rate holds for the second moments as suggested by our results for IPLA. The PGD variance attains the same rate but the overall variance of the estimates is significantly lower. While this can be seen as an advantage for PGD in convex settings, we believe this ``sticky'' behaviour is bad for non-convex optimisation. This noise is a feature of IPLA and can be beneficial to escape local minima in the non-convex setting.

\section{Conclusions}\label{sec:conclusion}

In this paper we propose a new algorithm to perform maximum marginal likelihood estimation  for latent variable models. The key observation of our work is that to obtain the maximum marginal likelihood estimator $
\bar{\theta}^\star$, it is sufficient to sample from~\eqref{eq:invariant_measure}. Our algorithm combines ideas coming from the literature on optimisation using Langevin based approaches with an interacting particle system targetting~\eqref{eq:invariant_measure} similar to that considered in \cite{kuntz2023particle}.

We exploit the connection with Langevin diffusions to provide convergence rates and nonasymptotic bounds which allow for careful tuning of the parameters needed to implement the algorithm (see Section~\ref{sec:bounds}).
We obtain our main results (Theorem~\ref{thm:main} and~\ref{thm:main stoch}) under rather strong assumptions, however, our proof techniques can be applied under weaker assumptions and we expect that similar results will hold for nonconvex (dissipative) $U$ and can be obtained adapting the techniques of \cite{zhang2023nonasymptotic}. Similarly, it is possible to relax the Lipschitz assumption on $U$ using the same framework.

While in many real world applications the strong convexity and Lipschitz assumptions do not hold, we believe that our present work constitutes the first building block towards developing and analysing this new class of algorithms, and paves the way to a new class of results which can be proved using a similar machinery.

The experiments in Section~\ref{sec:experiments} show that IPLA is competitive with other particle based methods (PGD; \cite{kuntz2023particle}) and Langevin based methods (SOUL; \cite{de2021efficient}) and significantly outperforms variational inference approaches.
The performances of IPLA and PGD are similar, as the latter is a modification of the former in which no noise is injected in the $\theta$-component. While this injection of noise might seem counterproductive, we found that it is crucial to obtain the explicit bounds in Theorem~\ref{thm:main} and~\ref{thm:main stoch} and we believe it will be beneficial when dealing with non-convex problems as suggested by the wide literature on this class of optimisation problems.
In addition, since for large values of $N$ the behaviour of IPLA and PGD gets closer, the theoretical results established in this work provide further validation and foundation for the work of \cite{kuntz2023particle}.

As observed in the introduction, any mechanism which samples from~\eqref{eq:invariant_measure} can be used to perform maximum marginal likelihood estimation. Our IPLA algorithm is one example but we anticipate that other techniques from the sampling literature could be employed to achieve the same task. For example, \cite{sharrock2023coinem} propose two variations of PGD based on Stein-variational gradient flows \cite{liu2016stein} and coin sampling \cite{sharrock2023coin}; we believe that similar ideas could be employed to provide equivalent variations of IPLA and we leave this for future work.


\begin{acknowledgement}
  This work has been supported by The Alan Turing Institute through the Theory and Methods Challenge Fortnights event ``Accelerating generative models and nonconvex optimisation'', which took place on 6-10 June 2022 and 5-9 Sep 2022 at The Alan Turing Institute headquarters. M.G. was supported by EPSRC grants EP/T000414/1, EP/R018413/2, EP/P020720/2, EP/R034710/1, EP/R004889/1, and a Royal Academy of Engineering Research Chair. F.R.C. and O.D.A. acknowledge support from the EPSRC (grant \#  EP/R034710/1). We thank Nikolas N\"usken, Juan Kuntz, and Valentin De Bortoli for fruitful discussions.
\end{acknowledgement}

\bibliographystyle{plain} 
\bibliography{sample}       
\newpage
\begin{appendix}
\section*{Appendix}
  \section{Proofs of Section~\ref{sec: analysis}}
  \label{app:ergodicity}
  
  Recall that we write continuous time processes such as \eqref{eq: IPS noised} in bold, their Euler-Maruyama discretisation as non-bold characters, and the continuous interpolation \eqref{eq: interpolation} of the discretisation as barred characters (so that $\bar{\theta}_{n \gamma}=\theta_n$ and $\bar{\theta}_t$ is an approximation of $\bm{\theta}^N_{ t}$). Finally, let us define, similarly to~\eqref{eq: rescaled cont.} the rescaled Euler-Maruyama discretisation 
  \begin{align}
  \label{eq:z_rescaling_intro}
  Z^N_n:=(\theta_n, N^{-1/2}X^{1,N}_n,...,N^{-1/2}X^{N,N}_n),
  \end{align} 
  the rescaled
  interpolated discretisation
  \begin{align}
  \bar{Z}^N_t &:=(\bar{\theta}_t, N^{-1/2}\bar{X}^{1,N}_t, \ldots, N^{-1/2}\bar{X}^{N,N}_t), \label{eq:Zt}
  \end{align}
  and finally the rescaled continuous process
  \begin{align}
  \bm{Z}^N_t &:=(\bm{\theta}^N_t, N^{-1/2}\bm{X}^{1,N}_t, \ldots, N^{-1/2}\bm{X}^{N, N}_t). \label{eq:Z_tN_cont}
  \end{align}
  We also denote the target measure of the rescaled system as $\pi_{\star,z}^N$. For notational convenience we define
  \begin{equation}
  \label{eq: Y defn}
  \bm{V}^{i,N}_t=(\bm{\theta}^N_t, \bm{X}^{i,N}_t), \;\;\;V^{i,N}_n=(\theta_n, X^{i,N}_n), \;\;\; \bar{V}^{i,N}_t=(\bar{\theta}_t, \bar{X}^{i,N}_t).
  \end{equation}
  Note that
  \begin{equation}
  \label{eq: V/Z identity}
  \frac{1}{N}\sum^N_{i=1} \Vert \bm{V}^{i,N}_t \Vert^2 = \Vert \bm{Z}^N_t \Vert^2,\;\;\; \frac{1}{N}\sum^N_{i=1} \Vert V^{i,N}_n \Vert^2 = \Vert Z^N_n \Vert^2, \;\;\; \frac{1}{N}\sum^N_{i=1} \Vert \bar{V}^{i,N}_t \Vert^2 = \Vert \bar{Z}^N_t \Vert^2,
  \end{equation}
  a fact that we shall use often in the following proofs.
  \subsection{Proof of Proposition~\ref{prop:e!}}
We can consider the system~\eqref{eq: IPS noised} as one SDE defined over $\mathbb{R}^d$ with $d=d_\theta+Nd_x$. The drift coefficient of this SDE is given by $b:\mathbb{R}^d\to\mathbb{R}^d$ with
\begin{equation}
\label{eq:fp_drift}
b(\theta, x_1,\dots, x_N):= (-N^{-1}\sum_{j=1}^N\nabla_\theta U(\theta, x_j), -\nabla_xU(\theta, x_1). \dots, -\nabla_xU(\theta, x_N)),
\end{equation}
and constant diffusion coefficient.
Under \Cref{Lipschitz assump}, it is easy to show that $b$ is Lipschitz continuous w.r.t. $\theta$ and $x_1, \dots, x_N$ and thus locally bounded, this and \Cref{Moments assump} ensure that there exists a unique strong solution to~\eqref{eq: IPS noised}
(see, for instance, \cite[Chapter 5, Theorem 2.9]{karatzas1991brownian}).
\subsection{Proof of Proposition~\ref{prop:invariant_measure}}
Recall that for a $\mathbb{R}^d$-valued SDE with drift $b:\mathbb{R}^d\to\mathbb{R}^d$ and diffusion coefficient $\sigma:\mathbb{R}^d \to\mathbb{R}^{d\times d}$, the `infinitesimal generator' $\mathcal{L}$ is the operator on $C^2(\mathbb{R}^d)$ given as
\begin{equation*}
\mathcal{L}= \langle b , \nabla \rangle + \frac{1}{2}\sigma \sigma^T : \nabla^2,
\end{equation*}
where $:$ denotes the Frobenius inner product and where one adopts the standard abuse of notation that treats operators like tensors, so that the first term for instance corresponds to the operator $\sum_{i=1}^d b_i \partial_{x_i}$. One has by \cite[1.17, page 696]{lelievre_stoltz_2016} that a probability measure $\pi$ on $\mathbb{R}^d$ is an invariant measure for the process with generator $\mathcal{L}$ if for any $\varphi \in C^\infty_c(\mathbb{R}^d)$ one has
\begin{equation*}
\int_{\mathbb{R}^d} \mathcal{L} \varphi \; d \pi=0.
\end{equation*}
Now let us consider \eqref{eq: IPS noised} as one SDE on $\mathbb{R}^{d_\theta+Nd_x}$ with generator $\mathcal{L}$. Then, for every $\varphi \in C^\infty_c(\mathbb{R}^{d_\theta+Nd_x})$ one has
\begin{align*}
\mathcal{L} \varphi &=-N^{-1}\sum_{j=1}^N \langle \nabla _\theta U(\theta, x_j), \nabla _\theta \varphi\rangle-\sum_{j=1}^N \langle \nabla _x U(\theta, x_j), \nabla _{x_j}\varphi \rangle+N^{-1}\Delta_\theta \varphi+\sum_{j=1}^N\Delta_{x_j}\varphi.
\end{align*}
Let $\rho:\mathbb{R}^{d_\theta+Nd_x} \to \mathbb{R}$ denote the density of $\pi^N_\star$, so that $\rho(\theta, \mathbf{x}_1,...,x_N)=ce^{-\sum^N_{i=1} U(\theta, x_i)}$ for some $c>0$, and observe that therefore by \Cref{Lipschitz assump} one has $\rho \in C^1(\mathbb{R}^{d_\theta+Nd_x})$. Now let us denote by $\divergence_\theta$ the divergence of a function as a function of $\theta$ alone, so that $\divergence_\theta(\varphi)=\frac{\varphi_1}{\partial \theta_1}+...+\frac{\varphi_{d_\theta}}{\partial \theta_{d_\theta}}$. Likewise one may define $\divergence_{x_j}$ for $j=1,..,N$. Then one has
\begin{equation*}
N^{-1}\divergence_\theta (\rho \nabla_\theta \varphi )=\biggr (-N^{-1}\sum_{j=1}^N \langle \nabla _\theta U(\theta, x_j), \nabla _\theta \varphi\rangle+N^{-1}\Delta_\theta \varphi \biggr)\rho
\end{equation*}
and 
\begin{equation*}
\divergence_{x_j}(\rho \nabla_{x_j} \varphi)=(-\langle \nabla _x U(\theta, x_j), \nabla _{x_j}\varphi \rangle+\Delta_{x_j}\varphi)\rho
\end{equation*}
so that since $d\pi = \rho \; d\theta d x_1 \ldots d x_N$ one has
\begin{align}\label{eq: int for inv}
\int_{\mathbb{R}^d} \mathcal{L} \varphi \; d\pi=N^{-1}\int \divergence_\theta (\rho \nabla_\theta \varphi ) \; d \theta d x_1 \ldots d x_N+\sum_{j=1}^N\int \divergence_{x_j}(\rho \nabla_{x_j} \varphi) \;  d \theta  d x_1 \ldots d x_N.
\end{align}
Now observe that, since $\varphi$ has compact support, the functions $\rho \nabla_{x_j} \varphi$ and $\rho \nabla_{x_j} \varphi$ for $j=1,2,..,N$ have compact support. Therefore one may apply the Divergence Theorem on a sufficiently large ball to conclude
\begin{equation*}
\int_{\mathbb{R}^{d_\theta}} \divergence_\theta (\rho \nabla_\theta \varphi ) \; d\theta=0,\;\;\; \int_{\mathbb{R}^{d_x}} \divergence_{x_j}(\rho \nabla_{x_j} \varphi) \;  d x_j =0
\end{equation*}
uniformly in all the coordinates not being integrated over. Then, since the integrals in \eqref{eq: int for inv} are finite, one may apply Fubini's Theorem to show that $\int_{\mathbb{R}^d} \mathcal{L} \varphi \; d\pi=0$.
\subsection{Proof of Proposition~\ref{prop:concentration}}
First we define the measure $\pi_\star^1$, i.e., the target measure for $N=1$. Observe that $\pi_\star^1$ is $\mu$-strongly log-concave, since $\pi_\star^1 \propto e^{-U(\theta, x)}$ and $U$ is strongly convex by~\Cref{Conv assump}. Therefore, by the Pr\'ekopa–-Leindler
inequality (see \cite[Theorem 7.1]{gardner2002brunn} for instance), $\pi^1_\Theta\propto e^{-K(\theta)} $ is $\mu$-strongly log-concave, i.e.,
\begin{align}
\label{eq:V_convex}
 \left\langle \theta - \theta', \nabla K(\theta) - \nabla K(\theta')\right\rangle \geq     \mu \|\theta - \theta'\|^2,
\end{align}
where $K(\theta) = -\log k(\theta)$ for all $\theta, \theta^\prime\in\Theta$, with $\mu$ given in~\Cref{Conv assump}.
Secondly, we note, since 
\begin{equation*}
\pi^N_\Theta (\theta) \propto \biggr(\int_{\mathbb{R}^{d_x}}e^{-U(\theta,x)} d x \biggr)^N,
\end{equation*}
one can write $\pi^N_\Theta \propto e^{-N K}$ as shown in \eqref{eq:mu_marginal_theta} and \eqref{eq:target_kappa}.

Next, we show that the Langevin SDE with drift $-\nabla K$ converges using the strong-convexity of $K$. It is well-known that the law of the overdamped Langevin SDE 
\begin{equation*}
d \bm{L}_t = -\nabla K(\bm{L}_t) d t+\sqrt{\frac{2}{N}}d \bm{B}_t, 
\end{equation*}
has invariant measure $\pi^N_\Theta$. Therefore, one may set the initial condition $\bm{L}_0$ to be a $\mathcal{F}_0$-measurable random variable satisfying $\mathcal{L}(\bm{L}_0)= \pi^N_\Theta$, and as a result $\mathcal{L}(\bm{L}_0)= \pi^N_\Theta$ for all $t>0$. Now using It\^{o }'s formula we may calculate
\begin{equation*}
\mathbb{E}\left[ \Vert \bm{L}_t - \bar{\theta}^\star \Vert^2 \right]= \mathbb{E}[\Vert \bm{L}_0 -  \bar{\theta}^\star \Vert^2]+\mathbb{E}\left[\int^t_0  2\langle -\nabla K(\bm{L}_s), \bm{L}_s - \bar{\theta}^\star \rangle d s\right]+ \frac{2d_\theta}{N}t,
\end{equation*}
which gives
\begin{equation*}
\frac{d}{d t}\mathbb{E}\left[ \Vert \bm{L}_t - \bar{\theta}^\star \Vert^2 \right]=  2 \mathbb{E}\left[ \langle - \nabla K(\bm{L}_t), \bm{L}_t - \bar{\theta}^\star \rangle\right]+\frac{2d_\theta}{N}.
\end{equation*}
Observing that $\bar{\theta}^\star$ is a minimiser of $K$, and so $\nabla K(\bar{\theta}^\star)=0$, we can apply~\eqref{eq:V_convex}, and obtain
\begin{align*}
\frac{d}{d t}\mathbb{E}\left[ \Vert \bm{L}_t - \bar{\theta}^\star \Vert^2 \right]&=  2\mathbb{E}\left[ \langle \nabla K(\bar{\theta}^\star) - \nabla K(\bm{L}_t), \bm{L}_t - \bar{\theta}^\star \rangle\right]+ \frac{2d_\theta}{N} \\
&\leq - 2 \mu \mathbb{E}\left[  \Vert \bm{L}_t - \bar{\theta}^\star \Vert^2 \right]+\frac{2d_\theta}{N}.
\end{align*}
Therefore
\begin{align*}
\frac{d}{d t}\mathbb{E}\left[ e^{2\mu t}\Vert \bm{L}_t - \bar{\theta}^\star \Vert^2 \right]\leq \frac{2d_\theta}{N} e^{2\mu t} .
\end{align*}
so that integrating both sides, by the initial condition and the Fundamental Theorem of Calculus one obtains
\begin{align*}
\mathbb{E}\left[ e^{2\mu t}\Vert \bm{L}_t - \bar{\theta}^\star \Vert^2 \right]\leq \mathbb{E} \left[\Vert \bm{L}_0 - \bar{\theta}^\star \Vert^2\right]+\frac{2d_\theta}{N \mu} e^{2\mu t} .
\end{align*}
Now, since $\mathcal{L}(\bm{L}_t)= \pi^N_\Theta$ for all $t>0$, one may divide through by $e^{2\mu t} $ to obtain that
\begin{equation*}
W_2(\pi^N_\Theta, \delta_{\bar{\theta}^\star} )^2 \leq \limsup_{t\to\infty} \mathbb{E}\left[ \Vert \bm{L}_t - \bar{\theta}^\star \Vert^2 \right] \leq \frac{2d_\theta}{N \mu},
\end{equation*}
as required.
  \subsection{Proof of Proposition~\ref{prop: conv to inv}}\label{sec:proof:conv_to_inv}
  \begin{proof}
  Let $v^\star= (\theta^\star, x^\star)$ be the minimiser of $U$, so $\nabla_\theta U(v^\star)$ and $\nabla_x U(v^\star)$ are vectors with identically zero entries. Let $z^\star=(\theta^\star, \frac{1}{\sqrt{N}}x^\star, ..., \frac{1}{\sqrt{N}}x^\star)$. We first bound $\mathbb{E}\left[\|\bm{Z}^N_t - z^\star\|^2\right]$ using It\^{o}'s formula
      \begin{align*}
  \mathbb{E}\left[\|\bm{Z}^N_t - z^\star\|^2 \right]&= \mathbb{E}\left[\|Z^N_0 - z^\star\|^2\right] - 2 \mathbb{E} \left[ \int_0^t \left\langle \bm{\theta}^N_s - \theta^\star, \frac{1}{N}\sum_{i=1}^N(\nabla_\theta U(\bm{V}^{i,N}_s) - \nabla_\theta U(v^\star))\right\rangle d s \right] \\
  &-2\frac{1}{N}\sum_{i=1}^N \mathbb{E}\left[ \int_0^t \left\langle \bm{X}^{i,N}_s - x^\star, \nabla _xU(\bm{V}^{i,N}_s) - \nabla_x U(v^\star)\right\rangle d s \right] +2 \frac{(N d_x + d_\theta)t}{N} \\
  & = \mathbb{E}\left[\|Z^N_0 - z^\star\|^2\right] - 2 \frac{1}{N}\sum_{i=1}^N\int_0^t \mathbb{E} \left[\left\langle \bm{V}^{i,N}_s - v^\star, \nabla U(\bm{V}^{i,N}_s) - \nabla U(v^\star)\right\rangle \right]d s\\
  &+ 2 \frac{(N d_x + d_\theta)t}{N}. 
  \end{align*}
  We compute derivatives of both sides and apply the convexity assumption \Cref{Conv assump} to obtain
  \begin{align*}
      \frac{d \mathbb{E}\left[\|\bm{Z}^N_t - z^\star\|^2\right]}{d t} &= -2 \frac{1}{N}\sum_{i=1}^N \mathbb{E} \left[\left\langle \bm{V}^{i,N}_t - v^\star, \nabla U(\bm{V}^{i,N}_t) - \nabla U(v^\star)\right\rangle \right]+ 2 \frac{N d_x + d_\theta}{N}\\
      &\leq -2 \mu \mathbb{E}\left[\|\bm{Z}^N_t - z^\star\|^2 \right]+ 2 \frac{N d_x + d_\theta}{N}.
  \end{align*}
  Consider next
  \begin{align*}
      \frac{d e^{2\mu t} \mathbb{E}\left[\|\bm{Z}^N_t - z^\star\|^2\right]}{d t} &= 2\mu e^{2 \mu t} \mathbb{E}\left[\|\bm{Z}^N_t - z^\star\|^2\right] + e^{2\mu t} \frac{d}{d t}\mathbb{E}\left[\|\bm{Z}^N_t - z^\star\|^2\right] \\
      &\leq 2\mu e^{2\mu t} \mathbb{E}\left[\|\bm{Z}^N_t - z^\star\|^2\right] + e^{2\mu t} \left(-2\mu \mathbb{E}\left[\|\bm{Z}^N_t - z^\star\|^2\right] + 2 \frac{N d_x + d_\theta}{N}\right) \\
      &= 2e^{2 \mu t}  \frac{N d_x + d_\theta}{N}.
  \end{align*}
  Integrating both sides gives
  \begin{align*}
      \mathbb{E}\left[\|\bm{Z}^N_t - z^\star\|^2 \right]e^{2 \mu t} - \mathbb{E} \left[\|Z_0^N - z^\star\|^2\right] \leq 2\frac{N d_x + d_\theta}{N} \int_0^t e^{2\mu s} d s.
  \end{align*}
  This implies that
  \begin{align}
  \label{eq:concentration1}
      \mathbb{E} \left[\|\bm{Z}^N_t - z^\star\|^2\right] \leq e^{-2\mu t} \mathbb{E}\left[ \|Z_0^N - z^\star\|^2 \right]+ \frac{N d_x + d_\theta}{N \mu}(1 - e^{-2\mu t}).
  \end{align}
  
Now we use the fact that $\pi_\star^N$ is the invariant measure of our SDE, i.e., $\pi_\star^N P_t^N = \pi_\star^N$ for the Markov kernel of our particle system $P_t^N$. It then follows by the definition \eqref{eq: rescaled cont.} that the rescaled system has a rescaled invariant measure, that we denote $\pi^N_{\star,z}$, so that if we denote the rescaled transition kernel of $\bm{Z}^N_t$ as $\tilde{P}_t^N$ then one has $\pi^N_{\star,z} \tilde{P}_t^N = \pi^N_{\star,z}$. Hence by \eqref{eq:concentration1} one can write
  \begin{align*}
      \int_{\mathbb{R}^{d_\theta+N d_x }} \|z- z^\star\|^2 \pi^N_{\star,z}(d z)&= \int_{\mathbb{R}^{d_\theta+N d_x }} \int_{\mathbb{R}^{d_\theta+N d_x }} \|z- z^\star\|^2 \tilde{P}_t^N(z', d z)\pi^N_{\star,z} (d z') \\
      &=\int_{\mathbb{R}^{d_\theta+N d_x }} \mathbb{E}^{z'}\left[\|\bm{Z}^N_t - z^\star\|^2\right] \pi^N_{\star,z}(d z'), \\
      &\leq e^{-2\mu t} \int_{\mathbb{R}^{d_\theta+N d_x }} \|z- z^\star\|^2 \pi^N_{\star,z}(d z)+ \frac{N d_x + d_\theta}{N \mu}(1 - e^{-2\mu t}),
  \end{align*}
  where $\mathbb{E}^{z'}$ denotes the expectation. w.r.t. $\tilde{P}_t^N$ starting at $z'$,
  which implies that
  \begin{align*}
      \int_{\mathbb{R}^{d_\theta+N d_x }} \|z- z^\star\|^2 \pi^N_{\star,z}(z)d z\leq \frac{d_\theta+N d_x}{N \mu}.
  \end{align*}
  We finalise this result by rewriting
  \begin{align}\label{eq:star_distance}
      \mathbb{E}\left[ \|\xi - z^\star\|^2 \right]^{1/2} \leq \left(\frac{d_\theta+N d_x}{N \mu}\right)^{1/2}
  \end{align}
       where $\xi$ is a $\mathcal{F}_0$-measurable (i.e. independent of the driving noise) random variable on $\mathbb{R}^{d_\theta+Nd_x}$  satisfying $\mathcal{L}(\xi)=\pi^N_{\star}$.
       
Now consider the IPS \eqref{eq: IPS noised} with initial condition $\xi$, which we shall denote $(\tilde{\bm{\theta}}_t, \tilde{\bm{X}}^{1,N}_t, ... , \tilde{\bm{X}}^{N,N}_t)$. Then since by Proposition~\ref{prop:invariant_measure} one has that $\pi^N_{\star}$ is an invariant measure for \eqref{eq: IPS noised}, one may conclude that the law of the system is equal to $\pi^N_\star$ for all time. Now consider the rescaling \eqref{eq: rescaled Z 2}, and denote by $\pi^N_{z,\star}$ the corresponding rescaled law of $\pi^N_\star$, so that in particular the rescaled initial condition $\xi_z:=(\xi_\theta, N^{-1/2}\xi_{x_1},..., N^{-1/2}\xi_{x_N})$ satisfies $\mathcal{L}(\xi_z) = \pi^N_{z,\star}$ and $\mathcal{L}(\tilde{\bm{Z}}^N_t)=\pi^N_{z,\star}$ for all $t>0$ (where $\tilde{\bm{Z}}^N_t$ is given in the same way as \eqref{eq: rescaled Z 2}). Then one calculates
      \begin{align*}
      \mathbb{E} \left[\|\bm{Z}^N_t - \tilde{\bm{Z}}_t^N\|^2\right] & = \mathbb{E} [\Vert Z^N_0-\xi_z \Vert^2]-
    \frac{2}{N} \sum^N_{i=1}\mathbb{E}\left[\int^t_0  \langle \nabla _\theta U(\bm{V}^{i,N}_s)-\nabla _\theta U(\tilde{\bm{V}}^{i,N}_s), \bm{\theta}^N_s-\tilde{\bm{\theta}}_s \rangle d s\right]\\
     &-   \frac{2}{N}\sum^N_{i=1} \mathbb{E}\left[\int^t_0 \langle \nabla _xU(\bm{V}^{i,N}_s)-\nabla _x U(\tilde{\bm{V}}^{i,N}_s), \bm{X}^{i,N}_s-\tilde{\bm{X}}^{i,N}_s \rangle d s \right]\\
     & =  \mathbb{E}[\Vert Z^N_0-\xi_z \Vert^2]- \frac{2}{N}\sum^N_{i=1} \mathbb{E}\left[\int^t_0 \langle \nabla U(\bm{V}^{i,N}_s)-\nabla  U(\tilde{\bm{V}}^{i,N}_s), \bm{V}^{i,N}_s-\tilde{\bm{V}}^{i,N}_s \rangle d s\right].
  \end{align*}
  Then we can differentiate, apply the convexity assumption~\Cref{Conv assump} to each term in the sum and then the identity \eqref{eq: V/Z identity} to obtain
  \begin{align*}
    \frac{d}{d t}  \mathbb{E}\left[\|\bm{Z}^N_t - \tilde{\bm{Z}}_t^N\|^2\right]  \leq -2\mu \mathbb{E}\left[\Vert \bm{Z}^N_s - \tilde{\bm{Z}}_s \Vert^2\right],
  \end{align*}
  so that 
  \begin{align*}
    \frac{d}{d t} \biggr ( e^{2\mu t} \mathbb{E}\left[\|\bm{Z}^N_t - \tilde{\bm{Z}}_t^N\|^2\right]  \biggr ) \leq 0,
  \end{align*}
  and therefore by the Fundamental Theorem of Calculus, since $\mathbb{E} [\|\bm{Z}^N_0 - \tilde{\bm{Z}}_0^N\|^2]= \mathbb{E}[ \Vert Z^N_0-\xi_z\Vert^2]$, integrating the above on $[0,t]$ and dividing by $e^{2\mu t}$ yields
  \begin{align*}
   \mathbb{E}\left[\|\bm{Z}^N_t - \tilde{\bm{Z}}_t^N\|^2\right]  \leq e^{-2\mu t}\mathbb{E}[\Vert Z^N_0-\xi_z \Vert^2] .
  \end{align*}
  Now note that by the definition of the Wassserstein distance, and since $\mathcal{L}(\tilde{\bm{\theta}}_t) = \pi^N_\Theta$, one may write
  \begin{align*}
  W_2(\mathcal{L}(\bm{\theta}^N_t), \pi^N_\Theta)^2\leq \mathbb{E}[ \Vert \bm{\theta}^N_t-\tilde{\bm{\theta}}_t \Vert^2] \leq \mathbb{E} [\Vert \bm{Z}^N_t-\tilde{\bm{Z}}^N_t \Vert^2] \leq e^{-2\mu t}\mathbb{E}[\Vert Z^N_0-\xi_z \Vert^2],
  \end{align*}
  which implies
  \begin{align}\label{eq:w2:bound}
  W_2(\mathcal{L}(\bm{\theta}^N_t), \pi^N_\Theta)\leq e^{-\mu t}\mathbb{E}[\Vert Z^N_0-\xi_z \Vert^2]^{1/2}.
  \end{align}
  Our strategy now is to bound $\mathbb{E}[\Vert Z^N_0-\xi_z \Vert^2]^{1/2}$ using a triangle inequality. Note that $\mathbb{E}[\Vert Z^N_0-\xi_z \Vert^2]^{1/2}$ is a norm, hence we can write
  \begin{align*}
      \mathbb{E}[\Vert Z^N_0-\xi_z \Vert^2]^{1/2} &\leq \mathbb{E}[\Vert Z^N_0-z^\star \Vert^2]^{1/2} + \mathbb{E}[\Vert z^\star-\xi_z \Vert^2]^{1/2}.
  \end{align*}
  Using \eqref{eq:star_distance} for the last term concludes the result.
  \end{proof}
  \subsection{Euler--Maruyama discretisation}\label{app:time_disc_1}
  \begin{rmrk}\label{rem: const C}
  In the following we let $C>0$ be a generic constant that changes from line to line, independent of $d_\theta, d_x, N$ and the stepsize $\gamma>0$.
  \end{rmrk}
  \subsubsection{Moment and increment bounds}
  \begin{lmm}\label{lem: numerics bdd moment}
  Let \Cref{Lipschitz assump},~\Cref{Conv assump}, \Cref{Moments assump} hold . Then for every $\gamma_0 \in (0, \min\{L^{-1}, 2\mu^{-1}\})$ there exists a constant $C>0$ independent of $t,n,N,\gamma, d_\theta, d_x$ such that for every $\gamma \in (0, \gamma_0)$, $n \in \mathbb{N}$,
  \begin{equation*}\label{eq: lem bound 1}
  \mathbb{E}[\lvert \lvert Z^N_n \rvert \rvert^2]\leq C(1+d_\theta/N+d_x).
  \end{equation*}
  \end{lmm}
  \begin{proof}
     Let $v^\star = (\theta^\star, x^\star)$ be the minimiser of the function $U$ ({where $\overline{\theta}^\star\not=\theta^\star$ in general}), and $z^\star=(\theta^\star, N^{-1/2}x^\star, ...., N^{-1/2}x^\star) \in \mathbb{R}^{d_\theta+Nd_x}$. Recall that a standard Gaussian on $\mathbb{R}^d$ has second moment equal to $d$, so that using the definition \eqref{eq: Y defn} of the process $V^{i,N}_t$ one has
  \begin{align*}
  \mathbb{E}\left[\Vert Z^N_{n+1} -z^\star\Vert^2 \right]&\leq \mathbb{E}\left[\Vert Z^N_n -z^\star\Vert^2\right] + \gamma^2 \mathbb{E}\left[\left(\frac{1}{N}\sum_{i=1}^N \Vert \nabla_{\theta} U(V^{i,N}_n) \Vert \right)^2 \right] + \frac{2\gamma d_\theta}{N}\notag\\
  &-\frac{2\gamma}{N}\sum_{j=1}^N \mathbb{E} \left[ \langle \theta_n-\theta^\star,  \nabla_{\theta} U(V_n^{j, N} )\rangle \right] \nonumber \\
  &+\frac{1}{N}\sum_{i=1}^N \biggr (  \gamma^2 \mathbb{E}\left[\Vert \nabla_x U(V_n^{i, N}) \Vert^2\right]+ 2\gamma d_x- 2\gamma \mathbb{E}\left[\langle X_n^{i, N}-x^\star, \nabla_x U(V_n^{i, N}) \rangle \right]\biggr ).
  \end{align*}
  Now note that 
  \begin{equation*}
  \frac{1}{N}\biggr(\sum_{i=1}^N\langle \theta, \nabla_\theta U(\theta, x_i) \rangle+ \langle x_i, \nabla_x U(\theta, x_i) \rangle \biggr) = \frac{1}{N}\sum_{i=1}^N \langle (\theta, x_i), \nabla U(\theta, x_i) \rangle,
  \end{equation*}
  and furthermore, by Young's inequality one has
  \begin{equation}\label{eq: sum identity}
  \biggr (\frac{1}{N}\sum_{i=1}^N a_i \biggr)^2 \leq \frac{1}{N}\sum_{i=1}^N a_i^2,
  \end{equation}
  which one may apply as well as the fact that 
  \begin{equation}\label{eq: pythag}
  \Vert \nabla U \Vert^2 =\Vert \nabla U_\theta \Vert^2+\Vert \nabla U_x \Vert^2 
  \end{equation}
  and $\nabla U(v^\star)=0$ to obtain that
  \begin{align}\label{eq: moment bound}
  \mathbb{E}\left[\Vert Z^N_{n+1} -z^\star\Vert^2 \right]&\leq  \mathbb{E}\left[\Vert Z^N_n -z^\star\Vert^2\right]- \frac{2\gamma}{N}\mathbb{E}\left[\sum_{i=1}^N \langle V^{i,N}_N-v^\star, \nabla U(V^{i,N}_n) -\nabla U(v^\star)\rangle\right] \nonumber \\
  &+ 2\gamma(\frac{d_\theta}{N}+ d_x)+\frac{\gamma^2}{N} \mathbb{E}\left[\biggr (\sum_{i=1}^N \Vert \nabla U(V^{i,N}_n) -\nabla U(v^\star)\Vert ^2\biggr )\right] .
  \end{align}
  Furthermore, since $U$ is convex, we may use co-coercivity (see Theorem 1 in \cite{gao2018properties}), given as
  \begin{equation*}
  \langle \nabla U(x)-\nabla U(y),x-y \rangle \geq \frac{1}{L}\Vert \nabla U(x)-\nabla U(y) \Vert^2,
  \end{equation*}
  for every $x,y \in \mathbb{R}^d$, to conclude for every $\gamma < \frac{1}{L}$ 
  \begin{equation*}
  - \frac{\gamma}{N}\mathbb{E}\left[\sum_{i=1}^N \langle V^{i,N}_N-v^\star, \nabla U(V^{i,N}_n) -\nabla U(v^\star)\rangle\right]+\frac{\gamma^2}{N}\mathbb{E}\left[\biggr (\sum_{i=1}^N \Vert \nabla U(V^{i,N}_n) -\nabla U(v^\star)\Vert ^2\biggr ) \right ] \leq 0.
  \end{equation*}
  Therefore by the convexity assumption \Cref{Conv assump} and \eqref{eq: V/Z identity}
  \begin{align*}
  \mathbb{E}\left[\Vert Z^N_{n+1} -z^\star\Vert^2 \right]&\leq  \mathbb{E}\left[\Vert Z^N_n -z^\star\Vert^2\right]- \frac{\gamma}{N}\mathbb{E}\left[\sum_{i=1}^N \langle V^{i,N}_N-v^\star, \nabla U(V^{i,N}_n) -\nabla U(v^\star)\rangle\right] + 2\gamma(\frac{d_\theta}{N}+d_x) \\
  &\leq  (1-\mu\gamma) \mathbb{E}\left[\Vert Z^N_n -z^\star\Vert^2\right]+ 2\gamma(\frac{d_\theta}{N}+d_x).
  \end{align*}
  Now recall that for $\alpha \in (-1,1)$, $\beta>0$ and a sequence $x_n \in [0,\infty)$ satisfying $x_{n+1}\leq\alpha x_n+\beta$ one has
  \begin{equation*}
      x_n \leq \alpha^n x_0+\beta \sum^n_{i=1}\alpha^{i-1} \leq x_0+\frac{\beta}{1-\alpha}.
  \end{equation*}
  Using this bound, for every $\gamma <\frac{2}{\mu}$ one obtains that
  \begin{align*}
  \mathbb{E}\left[\Vert Z^N_{n+1} -z^\star\Vert^2 \right]\leq  \mathbb{E} [\Vert Z^N_{0} -z^\star\Vert^2]+ \mu^{-1}(\frac{d_\theta}{N}+ d_x),
  \end{align*}
  and as a result, by \Cref{Moments assump} one has
  \begin{align*}
  \mathbb{E}\left[\Vert Z^N_{n} \Vert^2 \right]& \leq 2\mathbb{E}\left[\Vert Z_n^N -z^\star\Vert^2 \right]+2 \Vert z^\star\Vert^2 \nonumber \\
  &\leq C(1+ \frac{d_\theta}{N}+ d_x),
  \end{align*}
  where $C>0$ is the generic constant introduced in Remark \ref{rem: const C}.
  \end{proof}
  \begin{lmm}\label{lem: numerics bdds}
  Let \Cref{Lipschitz assump},~\Cref{Conv assump}, \Cref{Moments assump} hold and consider the interpolated discretisation \eqref{eq:Zt}. Then for every $\gamma_0 \in (0, \min\{L^{-1}, 2\mu^{-1}\})$ there exists a constant $C>0$ independent of $t,n,N,\gamma, d_\theta, d_x$ such that for every $\gamma \in (0, \gamma_0)$, $n \in \mathbb{N}$, $t \geq 0$
  \begin{equation*}
      \mathbb{E}[\lvert \lvert \bar{Z}^N_t-\bar{Z}^N_{\kappa_\gamma(t)} \rvert \rvert^2]\leq C(1+d_\theta/N+d_x)\gamma.
  \end{equation*}
  and
  \begin{equation}\label{eq: lem bound 3}
      \mathbb{E}[\lvert \lvert \bm{Z}^N_t-\bm{Z}^N_{\kappa_\gamma(t)}  \rvert \rvert^2]\leq C(1+d_\theta/N+d_x)\gamma.
  \end{equation}
  
  \end{lmm}
  \begin{proof}
  Recall that $\kappa_\gamma(t)=\gamma\lfloor \gamma^{-1} t \rfloor$ is the projection back onto the grid $\{0,\gamma, 2\gamma,..\}$. Using the interpolation $\bar{Z}^N_t$ of $Z^N_n$ given in \eqref{eq:z_rescaling_intro}, given that $\bar{Z}^N_{\kappa_\gamma(t)} = Z^N_{\kappa_\gamma(t)/\gamma}$ one calculates by \eqref{eq: V/Z identity} and \eqref{eq: pythag}, and since the second moment of a standard $d$-dimensional Gaussian is $d$, that
  \begin{align*}
  \mathbb{E}\left[\Vert \bar{Z}^N_t -\bar{Z}^N_{\kappa_\gamma(t)}  \Vert^2 \right]\leq  2\lvert t-\kappa_\gamma(t) \rvert(\frac{d_\theta}{N}+d_x)+\frac{\lvert t-\kappa_\gamma(t) \rvert^2}{N}  \mathbb{E}\left[\sum_{i=1}^N \Vert \nabla U(V^{i,N}_{\kappa_\gamma(t)/\gamma}) 
  \Vert ^2\right] ,
  \end{align*}
  so that applying the Lipschitz assumption \Cref{Lipschitz assump} and \eqref{eq: V/Z identity} one has, since $\lvert t-\kappa_\gamma(t) \rvert\leq 1$,
  \begin{align*}
  \mathbb{E}\left[\Vert \bar{Z}^N_t -\bar{Z}^N_{\kappa_\gamma(t)} \Vert^2 \right]&\leq  C\lvert t-\kappa_\gamma(t) \rvert(d_\theta/N +d_x+\frac{1}{N}\sum_{i=1}^N \mathbb{E}[\Vert V^{i,N}_{\kappa_\gamma(t)/\gamma} \Vert^2]) \leq C\gamma(1+\mathbb{E}[\Vert Z^N_{\kappa_\gamma(t)/\gamma} \Vert^2]),
  \end{align*}
  and therefore that the first stated result follows from Lemma \ref{lem: numerics bdd moment}. We shall use a similar argument to prove \eqref{eq: lem bound 3}. In particular, we have by \eqref{eq:concentration1} and \Cref{Moments assump} that 
  \begin{equation*}
  \mathbb{E}[\Vert \bm{Z}^N_t \Vert^2]\leq C(1+ d_\theta/N + d_x),
  \end{equation*}
  so by It\^{o }'s formula
  \begin{align*}
  \mathbb{E}[\Vert \bm{Z}^N_t- \bm{Z}^N_{\kappa_\gamma(t)}\Vert^2]&=-\mathbb{E}\biggr [\int^t_{\kappa_\gamma(t)} \biggr ( \frac{2}{N}\sum_{i=1}^N\langle \nabla_\theta U(\bm{V}^{i,N}_u), \bm{\theta}^N_u \rangle +\frac{2}{N}\sum_{i=1}^N \langle \nabla_x U(\bm{V}^{i,N}_u), \bm{X}^{i,N}_u \rangle + 2(d_\theta/N+d_x) \biggr ) d u \biggr]  \\
  & =\int^t_{\kappa_\gamma(t)} \biggr ( \frac{2}{N}\sum_{i=1}^N\mathbb{E}[\langle \nabla U(\bm{V}^{i,N}_u), \bm{V}^{i,N}_u \rangle] + 2(d_\theta/N+d_x) \biggr ) d u,
  \end{align*}
  and therefore by Cauchy-Schwarz, Young's inequality, the Lipschitz Assumption \Cref{Lipschitz assump} and \eqref{eq: V/Z identity} one has
  \begin{align*}
  \mathbb{E}[\Vert \bm{Z}^N_t- \bm{Z}^N_{\kappa_\gamma(t)}\Vert^2]&\leq C\int^t_{\kappa_\gamma(t)} \biggr ( \frac{1}{N}\sum_{i=1}^N(1+\mathbb{E}[\Vert \bm{V}^{i,N}_u \Vert^2]) + d_\theta/N+d_x \biggr ) d u  \\
  & \leq C\int^t_{\kappa_\gamma(t)} \biggr ( 1+\mathbb{E}[\Vert \bm{Z}^N_u \Vert^2] + d_\theta/N+d_x \biggr ) d u  \\
  & \leq C(1+ d_\theta/N + d_x)\lvert t-\kappa_\gamma(t)\rvert,
  \end{align*}
  where $C>0$ is the generic constant introduced in Remark \ref{rem: const C}.
\end{proof}
  
  \subsubsection{Proof of Proposition~\ref{prop: discr error 1}}
  \begin{proof}
  In order to prove~\eqref{eq: Lipschtiz disc error} we consider the convergence of the entire rescaled discretised IPS $Z^N_n$ to the original continuous rescaled process $\bm{Z}^N_t$. In particular, noting that $Z^N_n$ is an approximation of $\bm{Z}^N_{n\gamma}$, we shall prove that
  \begin{equation} \label{eq: Lipschitz disc error sys}
  \mathbb{E}\left[\Vert Z^N_n -\bm{Z}^N_{n\gamma} \Vert^2\right] \leq C(1+ d_\theta/N + d_x) \gamma,
  \end{equation}
  at which point~\eqref{eq: Lipschtiz disc error} follows immediately. The interpolated process $\bar{Z}^N_t$ given as
  \begin{align*}
  \bar{Z}^N_t:=(\bar{\theta}_t, N^{-1/2}\bar{X}^{i,N}_t,...,N^{-1/2}\bar{X}^{i,N}_t).
  \end{align*}
  via the interpolated discretisation \eqref{eq: interpolation} satisfies $\bar{Z}^N_{\gamma n} = Z^N_n$, so we shall actually prove
  \begin{equation} \label{eq: Lipschitz disc error sys 2}
  \mathbb{E}\left[\Vert \bar{Z}^N_t -\bm{Z}^N_{t} \Vert^2\right] \leq C(1+ d_\theta/N + d_x) \gamma.
  \end{equation}
  Since $\bar{Z}^N_t$ is an It\^o process, so we can apply It\^o's formula for $x \mapsto \lvert x \rvert^2$ to obtain
  \begin{align*}
  \lvert\lvert \bar{Z}^N_t - \bm{Z}^N_{t} \rvert\rvert^2 &= -\frac{2}{N} \sum_{i=1}^N  \int^t_0 \langle \nabla_{\theta} U(\bar{\theta}_{\kappa_\gamma(s)},\bar{X}_{\kappa_\gamma(s)}^{i, N}) - \nabla_{\theta} U(\bm{\theta}^N_{s}, \bm{X}^{i,N}_{s}), \bar{\theta}_s - \bm{\theta}^N_s \rangle d s \nonumber \\ 
  & - \frac{2}{N} \sum_{i=1}^N\int^t_0 \langle \nabla_x U(\bar{\theta}_{\kappa_\gamma(s)}, \bar{X}_{\kappa_\gamma(s)}^{i, N}) - \nabla_x U(\bm{\theta}^N_{s}, \bm{X}^{i,N}_{s}), \bar{X}^{i,N}_s - \bm{X}^{i,N}_{s} \rangle d s  \\ 
  &= -\frac{2}{N} \sum_{i=1}^N  \int^t_0 \langle \nabla U(\bar{V}^{i,N}_{\kappa_\gamma(s)}) - \nabla U(\bm{V}_{s}), \bar{V}^{i,N}_s - \bm{V}^{i,N}_{s} \rangle d s ,
  \end{align*}
  where $\bar{V}^{i,N}_t=(\bar{\theta}_t, \bar{X}^{i,N}_t)$ and $\bm{V}^{i,N}_t=(\bm{\theta}^N_t, \bm{X}^{i,N}_t)$ for $i=1,2,...,N$ and $t>0$. For convenience let us define 
  \begin{equation*}
  e^{i,N}_t:= \bar{V}^{i,N}_t - \bm{V}^{i,N}_t, \;\; e_t = \bar{Z}^N_t - \bm{Z}^N_t, 
  \end{equation*}
  so that in particular
  \begin{equation}\label{eq: ei identity}
  \frac{1}{N}\sum^N_{i=1} \Vert e^{i,N}_t \Vert^2 = \Vert e_t \Vert^2. 
  \end{equation}
  Using the above, one may apply It\^o's formula again for $x \mapsto e^{\mu t }\lvert x \rvert^2$ to obtain
  \begin{equation}\label{eq: newdif}
   e^{\mu t } \Vert e_t \Vert^2 = \int^t_0 \mu e^{\mu s} \Vert e_s \Vert^2+  2\int^t_0 e^{\mu s }\biggr (- \frac{1}{N} \sum_{i=1}^N \langle \nabla U(\bar{V}^{i,N}_{\kappa_\gamma(s)}) - \nabla U(\bm{V}_{s}), e^{i,N}_s \rangle \biggr) d s.
  \end{equation}
  Let us now bound $-N^{-1}\sum_{i=1}^N \langle \nabla U(\bar{V}^{i,N}_{\kappa_\gamma(s)}) - \nabla U(\bm{V}_{s}), e^{i,N}_s \rangle$ by splitting as
  \begin{align}\label{eq: dif1}
-\frac{1}{N} \sum_{i=1}^N \langle \nabla U(\bar{V}^{i,N}_{\kappa_\gamma(s)}) - \nabla U(\bm{V}_{s}), e^{i,N}_s \rangle &= -\frac{1}{N}\sum_{i=1}^N    \langle \nabla U(\bar{V}^{i,N}_s) - \nabla U(\bm{V}_s^{i,N}), e^{i,N}_s \rangle  \nonumber \\ 
  & -\frac{1}{N}\sum_{i=1}^N \langle \nabla U(\bar{V}^{i,N}_{\kappa_\gamma(s)}) - \nabla U(\bar{V}^{i,N}_s), e^{i,N}_s \rangle \nonumber \\
  &:=r_1(s)+r_2(s)
  \end{align}
We then control the first term in the same way as in the proof of Proposition~\ref{prop:concentration}, and the discretisation error $r_2(s)$ using standard bounds for stochastic processes. Specifically, for the former we have by the convexity assumption~\Cref{Conv assump} and a variation on \eqref{eq: V/Z identity} that
  \begin{align}\label{eq: e1}
  r_1(s)&= -\frac{1}{N} \sum_{i=1}^N \langle \nabla U(\bar{V}^{i,N}_s) - \nabla U(\bm{V}_s), e^{i,N}_s \rangle  \nonumber \\
  & \leq -\mu \frac{1}{N} \sum_{i=1}^N  \Vert V_s^{i, N} - \bm{V}^{i,N}_s\Vert^2   \nonumber \\
  & \leq -\mu  \Vert e_s \Vert^2 .
  \end{align}
  For the second part, the Cauchy-Schwarz inequality, Young's inequality and~\Cref{Lipschitz assump} give
  \begin{align}\label{eq: e2}
  r_2(s)& = \frac{1}{N}\sum_{i=1}^N \langle \nabla U(\bar{V}^{i,N}_s)-\nabla U(\bar{V}^{i,N}_{\kappa_\gamma(s)}) , e^{i,N}_s \rangle \nonumber \\
  &\leq \frac{1}{N}\sum_{i=1}^N \Vert \nabla U(\bar{V}^{i,N}_s)-\nabla U(\bar{V}^{i,N}_{\kappa_\gamma(s)}) \Vert \cdot \Vert e^{i,N}_s \Vert \nonumber \\
  &\leq \frac{1}{N}\sum_{i=1}^N  (\frac{1}{2\mu}\Vert \nabla U(\bar{V}^{i,N}_{\kappa_\gamma(s)}) - \nabla U(\bar{V}^{i,N}_s) \rvert\rvert^2 + \frac{\mu}{2}\Vert e^{i,N}_s \Vert^2 ) \nonumber \\
  & \leq  C \Vert \bar{Z}^N_{\kappa_\gamma(s)} - \bar{Z}^N_s \Vert^2 +\frac{\mu}{2} \Vert e_s\Vert^2 . 
  \end{align}
  Therefore, by Lemma \ref{lem: numerics bdds}, combining~\eqref{eq: e1} and~\eqref{eq: e2} into~\eqref{eq: dif1} and taking expectation one obtains
   \begin{align}
-\frac{1}{N} \sum_{i=1}^N \mathbb{E} [\langle \nabla U(\bar{V}^{i,N}_{\kappa_\gamma(s)}) - \nabla U(\bm{V}_{s}), e^{i,N}_s \rangle ]\leq C(1+ d_\theta/N + d_x)\gamma -\frac{\mu}{2}\mathbb{E}[\Vert e_s\Vert^2],
  \end{align}
  which one substitutes into \eqref{eq: newdif} to obtain
    \begin{equation}
   e^{\mu t }\mathbb{E}[ \Vert e_t \Vert^2 ]\leq  \int^t_0 C\gamma (1+ d_\theta/N + d_x)e^{\mu s} d s \leq C(1+ d_\theta/N + d_x)\gamma e^{\mu t },
  \end{equation}
  at which point the result follows by dividing through by $e^{\mu t }$.
  \end{proof}
  
  \subsubsection{Proof of Proposition~\ref{prop: discr error 2}}
  \begin{proof}
  We follow a similar strategy to the proof of Proposition \ref{prop: discr error 1}, but using a slightly more delicate strategy to bound $r_2$, defined in \eqref{eq: dif1}. In particular, note that if we denote
  \begin{equation*}
  \zeta^{i,N}_s:= \sqrt{2}(N^{-1/2}(\bm{B}^{0,N}_s-\bm{B}^{0,N}_{\kappa_\gamma(s)}), \bm{B}^{i,N}_s-\bm{B}^{i,N}_{\kappa_\gamma(s)}) 
  \end{equation*}
  then
  \begin{equation*}
  \bar{V}^{i,N}_s=\bar{V}^{i,N}_{\kappa_\gamma(s)}-(s-\kappa_\gamma(s)) \biggr (\frac{1}{N}\sum^N_{j=1} \nabla_\theta U(\bar{V}^{j,N}_{\kappa_\gamma(s)}), \;\nabla_x U(\bar{V}^{j,N}_{\kappa_\gamma(s)}) \biggr)+\zeta^{i,N}_s,
  \end{equation*}
  so we can split $r_2$ as
  \begin{align*}
  r_2(s)& = \frac{1}{N}\sum_{i=1}^N \langle \nabla U(\bar{V}^{i,N}_s)-\nabla U(\bar{V}^{i,N}_{\kappa_\gamma(s)}) , e^{i,N}_s \rangle \nonumber \\ 
  &= \frac{1}{N}\sum_{i=1}^N \langle \nabla U(\bar{V}^{i,N}_s)-\nabla U(\bar{V}^{i,N}_{\kappa_\gamma(s)}+\zeta^{i,N}_s) , e^{i,N}_s \rangle \nonumber \\
  & +\frac{1}{N}\sum_{i=1}^N \langle \nabla U(\bar{V}^{i,N}_{\kappa_\gamma(s)}+\zeta^{i,N}_s)-\nabla U(\bar{V}^{i,N}_{\kappa_\gamma(s)}) , e^{i,N}_s \rangle  \nonumber \\
  &:= r_{2,1}(s)+r_{2,2}(s).
  \end{align*}
 Firstly let us bound $r_{2,1}$. By Cauchy-Schwarz, Young's inequality and \eqref{eq: ei identity} one has
  \begin{align*}
  r_{2,1}(s) &\leq \frac{1}{N}\sum_{i=1}^N (C\Vert \bar{V}^{i,N}_s - \bar{V}^{i,N}_{\kappa_\gamma(s)}-\zeta^{i,N}_s \Vert^2 +\frac{\mu}{6}\Vert e^{i,N}_s \Vert^2) \\
  & \leq \gamma^2 \frac{1}{N}\sum_{i=1}^N C\biggr (\biggr\Vert \frac{1}{N}\sum_{j=1}^N  \nabla _\theta U(\bar{V}^{j,N}_{\kappa_\gamma(s)}) \biggr \Vert^2+  \Vert \nabla_x U(\bar{V}^{i,N}_{\kappa_\gamma(s)})\Vert^2     \biggr)+\frac{\mu}{6}\Vert e_t \Vert^2.
  \end{align*}
  Now one may apply \eqref{eq: sum identity} to terms arising from the sum in the first term above, so that by the Lipschitz assumption and Lemma \ref{lem: numerics bdd moment}
  \begin{align}\label{eq: k21}
   \mathbb{E} [r_{2,1}(s)] &\leq \gamma^2 \frac{C}{N}\sum_{i=1}^N \mathbb{E} [\Vert \nabla U(\bar{V}^{i,N}_{\kappa_\gamma(s)} ) \Vert^2] +\frac{\mu}{6}\mathbb{E} [\Vert e_s \Vert^2] \nonumber \\
  & \leq C(1+d_\theta/N +d_x)\gamma^2 +\frac{\mu}{6}\mathbb{E} [\Vert e_s \Vert^2].
  \end{align}
  Now we shall need to split $r_{2,2}$ as follows
  \begin{align*}
  r_{2,2}(s) &= \frac{1}{N}\sum_{i=1}^N \langle \nabla U(\bar{V}^{i,N}_{\kappa_\gamma(s)}+\zeta^{i,N}_s)-\nabla U(\bar{V}^{i,N}_{\kappa_\gamma(s)}) , e^{i,N}_s \rangle \nonumber \\ 
  &=\frac{1}{N}\sum_{i=1}^N \langle \nabla U(\bar{V}^{i,N}_{\kappa_\gamma(s)} +\zeta^{i,N}_s)-\nabla U(\bar{V}^{i,N}_{\kappa_\gamma(s)}) -\Delta U(\bar{V}^{i,N}_{\kappa_\gamma(s)})\zeta^{i,N}_s, e^{i,N}_s\rangle \nonumber \\
  &+\frac{1}{N}\sum_{i=1}^N \langle \Delta U(\bar{V}^{i,N}_{\kappa_\gamma(s)})\zeta^{i,N}_s, e^{i,N}_s-e^{i,N}_{\kappa_\gamma(s)}\rangle\\
  &+\frac{1}{N}\sum_{i=1}^N\langle \Delta U(\bar{V}^{i,N}_{\kappa_\gamma(s)})\zeta^{i,N}_s, e^{i,N}_{\kappa_\gamma(s)}\rangle \nonumber \\
  &:=r_{2,2,1}(s)+r_{2,2,2}(s)+r_{2,2,3}(s).
  \end{align*}
  For the first term in this splitting we use the smoothness assumption \Cref{Smoothness assump} and \cite[Lemma 5]{kumar2019milstein} with $\gamma=0$ to obtain by the definition of $\zeta^{i,N}_t$ (since a standard Gaussian on $\mathbb{R}^d$ has fourth moment equal to $Cd^2$) that
  \begin{align*}
  \mathbb{E} [r_{2,2,1}(s)] & \leq \frac{1}{N}\sum_{i=1}^N (C \mathbb{E} [\Vert \nabla U(\bar{V}^{i,N}_{\kappa_\gamma(t)}+\zeta^{i,N}_s)-\nabla U(\bar{V}^{i,N}_{\kappa_\gamma(s)}) -\Delta U(\bar{V}^{i,N}_{\kappa_\gamma(s)})\zeta^{i,N}_s \Vert^2]+\frac{\mu}{6} \mathbb{E} [\Vert e^{i,N}_s \Vert^2]) \nonumber \\
  & \leq C\frac{1}{N}\sum_{i=1}^N \mathbb{E} [\Vert \zeta^{i,N}_s \Vert^4] +\frac{\mu}{6}\mathbb{E} [\Vert e_t \Vert^2] 
  \end{align*}
  Now note that by definition, since a standard Gaussian $\zeta$ on $\mathbb{R}^d$ satisfies $\mathbb{E}[\Vert \zeta \Vert^2]=d$, $\mathbb{E}[\Vert \zeta \Vert^4]=Cd^2$, one has
  \begin{equation*}
  \mathbb{E} [\Vert \zeta^{i,N}_s \Vert^4]=C(N^{-2}d_\theta^2+d_x^2+N^{-1}d_\theta d_x)\gamma^2\leq C(d_\theta/N+d_x)^2\gamma^2,
  \end{equation*}
  so that
  \begin{equation}
   \label{eq: k221}
  \mathbb{E} [r_{2,2,1}(s)]\leq C (N^{-1}d_\theta+d_x)^2\gamma^2 +\frac{\mu}{6}\mathbb{E} [\Vert e_s \Vert^2] .
  \end{equation}
  To control $r_{2,2,2}$, first by Young's inequality (recalling that the expression in brackets below is an element of $\mathbb{R}^{d_\theta+d_x}$ and not an inner product) we have the bound
  \begin{align*} 
  &\Vert \Delta U(\bar{V}^{i,N}_{\kappa_\gamma(s)})\zeta^{i,N}_t \Vert \cdot \Vert e^{i,N}_s-e^{i,N}_{\kappa_\gamma(s)} \Vert \\
  &\leq  \Vert \Delta U(\bar{V}^{i,N}_{\kappa_\gamma(t)})\zeta^{i,N}_s \Vert \cdot \int^s_{\kappa_\gamma(s)}\biggr \Vert \biggr (\frac{1}{N}\sum_{j=1}^N(\nabla_\theta U(\bm{V}_u^{j,N})-\nabla _\theta U(\bar{V}_{\kappa_\gamma(s)}^{j,N})), \nabla_x U(\bm{V}_u^{i,N})-\nabla_xU(\bar{V}_{\kappa_\gamma(s)}^{i,N})  \biggr) \biggr \Vert d u  \\
  & \leq   \int^s_{\kappa_\gamma(s)}\biggr (C\Vert \Delta U(\bar{V}^{i,N}_{\kappa_\gamma(s)})\zeta^{i,N}_s \Vert ^2+ \frac{\mu}{18L}\biggr \Vert \frac{1}{N}\sum_{j=1}^N(\nabla_\theta U(\bm{V}_u^{j,N})-\nabla _\theta U(\bar{V}_{\kappa_\gamma(s)}^{j,N})) \biggr \Vert^2 \biggr ) d u\nonumber \\
  &+\int^s_{\kappa_\gamma(s)}\frac{\mu}{18L}\Vert \nabla_x U(\bm{V}_u^{i,N})-\nabla_xU(\bar{V}_{\kappa_\gamma(s)}^{i,N}) \Vert^2 d u .
  \end{align*}
  Now note that by the independence of $\zeta^{i,N}_t$ from $\bar{V}^{i,N}_{\kappa_\gamma(t)}$ and \Cref{Smoothness assump} one has
  \begin{align*}
  \mathbb{E}[\Vert \Delta U(\bar{V}^{i,N}_{\kappa_\gamma(s)})\zeta^{i,N}_s \Vert ^2]& \leq \mathbb{E}[\Vert \Delta U(\bar{V}^{i,N}_{\kappa_\gamma(s)}) \Vert^2] \mathbb{E} [\Vert \zeta^{i,N}_s \Vert ^2] \nonumber \leq C \mathbb{E} [\Vert \bar{V}^{i,N}_{\kappa_\gamma(s)} \Vert^2] (1+d_\theta /N+d_x)\gamma,
  \end{align*}
  so that by \eqref{eq: sum identity}
  \begin{align*} 
  &\mathbb{E} [\Vert \Delta U(\bar{V}^{i,N}_{\kappa_\gamma(t)})\zeta^{i,N}_s \Vert \cdot\Vert e^{i,N}_s-e^{i,N}_{\kappa_\gamma(s)} \Vert ]\nonumber \\
  &\leq  \int^s_{\kappa_\gamma(s)}C\gamma(1+d_\theta /N+d_x) \mathbb{E} [\Vert \bar{V}^{i,N}_{\kappa_\gamma(s)} \Vert^2] + \frac{\mu}{18NL}\sum_{j=1}^N \mathbb{E}[\Vert \nabla_\theta U(\bm{V}_u^{j,N})-\nabla _\theta U(\bar{V}_{\kappa_\gamma(s)}^{j,N}) \Vert^2] d u \nonumber \\
  &+\int^s_{\kappa_\gamma(s)} \frac{\mu}{12 }\mathbb{E}[\Vert \nabla_x U(\bm{V}_u^{i,N})-\nabla_xU(\bar{V}_{\kappa_\gamma(s)}^{i,N}) \Vert^2] d u ,
  \end{align*}
  and therefore by \Cref{Lipschitz assump},  Lemma~\ref{eq: lem bound 1} (recalling $\bar{Z}^N_{\kappa_\gamma(t)} =Z^N_{\kappa_\gamma(t)/\gamma}$) , and since $\lvert t - \kappa_\gamma(t)\rvert \leq \gamma$, multiplying by $N^{-1}$ and summing over $i$ one has
  \begin{align*} 
  \mathbb{E}[r_{2,2,2}(s)]&:=\frac{1}{N}\sum_{i=1}^N \mathbb{E}[ \Vert \Delta U(\bar{V}^{i,N}_{\kappa_\gamma(s)})\zeta^{i,N}_s \Vert \cdot\Vert e^{i,N}_s-e^{i,N}_{\kappa_\gamma(s)} \Vert ] \\
  &\leq  \int^s_{\kappa_\gamma(s)}C\gamma(1+d_\theta /N+d_x) \mathbb{E} [\Vert \bar{Z}^{i,N}_{\kappa_\gamma(s)} \Vert^2]+ \frac{\mu}{18NL}\sum_{j=1}^N\int^s_{\kappa_\gamma(s)} \mathbb{E}[ \Vert \nabla U(\bm{V}_u^{j,N})-\nabla U(\bar{V}_{\kappa_\gamma(s)}^{j,N}) \Vert^2 ]d u , \\
  &\leq  C \gamma^2(1+d_\theta/N+d_x)+ \frac{\mu}{18NL}\sum_{j=1}^N \int^s_{\kappa_\gamma(s)} \mathbb{E}[ \Vert \nabla U(\bm{V}_u^{j,N})-\nabla U(\bar{V}_{\kappa_\gamma(s)}^{j,N}) \Vert^2 ]d u ,
  \end{align*}
  Now one has by \Cref{Lipschitz assump}
  \begin{align*}
  \mathbb{E}[ \Vert \nabla U(\bm{V}_u^{j,N})-\nabla U(\bar{V}_{\kappa_\gamma(s)}^{j,N}) \Vert^2 ] &\leq L \mathbb{E}[ \Vert \bm{V}_u^{j,N}-\bar{V}_{\kappa_\gamma(s)}^{j,N} \Vert^2 ] \nonumber \\
  &\leq 3L\mathbb{E}[\Vert \bm{V}_u^{j,N}-\bm{V}_s^{j,N} \Vert^2]+3L\mathbb{E}[\Vert e^{j,N}_s\Vert^2]+3L\mathbb{E}[\Vert \bar{V}_s^{j,N}-\bar{V}_{\kappa_\gamma(u)}^{j,N} \Vert^2]
  \end{align*}
  so that by \eqref{eq: V/Z identity} and Lemma \ref{lem: numerics bdds}, and since $\gamma \in [0,1]$
  \begin{align}\label{eq: k222}
  \mathbb{E}[r_{2,2,2}(s)]&\leq  C \gamma^2(1+d_\theta/N+d_x)+ \frac{\mu}{6 }\int^s_{\kappa_\gamma(s)}\mathbb{E}[\Vert e_s\Vert^2]d u \nonumber\\
  &+\frac{\mu}{6 N}\sum_{j=1}^N\int^s_{\kappa_\gamma(s)}  (\mathbb{E}[\Vert \bm{V}_u^{j,N}-\bm{V}_s^{j,N} \Vert^2]+\mathbb{E}[\Vert \bar{V}_s^{j,N}-\bar{V}_{\kappa_\gamma(s)}^{j,N} \Vert^2]) d u\nonumber \\
  &\leq   \gamma^2(1+d_\theta/N+d_x)+ \frac{\mu}{6 }\int^s_{\kappa_\gamma(s)}\mathbb{E}[\Vert e_s\Vert^2]du+\frac{\mu}{6 }\int^s_{\kappa_\gamma(s)} (\mathbb{E}[\Vert \bm{Z}_u^N-\bm{Z}_s^N \Vert^2]+\mathbb{E}[\Vert \bar{Z}_s^N-\bar{Z}_{\kappa_\gamma(s)}^N \Vert^2 ])d u  \nonumber \\
  &\leq  \gamma^2(1+d_\theta/N+d_x)+\frac{\mu}{6} \mathbb{E}[\Vert e_s\Vert^2].
  \end{align}
  Finally since $\zeta^{i,N}_s$ is independent of $\bar{V}^{i,N}_{\kappa_\gamma(s)}$ and $e^{i,N}_{\kappa_\gamma(s)}$ we have
  \begin{equation}\label{eq: k223}
  \mathbb{E}\left[r_{2,2,3}(s)\right]=\frac{1}{N}\sum_{i=1}^N\mathbb{E}\left[\langle \Delta U(\bar{V}^{i,N}_{\kappa_\gamma(s)})\zeta^{i,N}_s, e^{i,N}_{\kappa_\gamma(s)}\rangle\right] =0.
  \end{equation}
  Therefore, substituting ~\eqref{eq: e1}
  %
  , ~\eqref{eq: k21}, \eqref{eq: k221}, \eqref{eq: k222} and \eqref{eq: k223} into~\eqref{eq: dif1}, one has
   \begin{align}
-\frac{1}{N} \sum_{i=1}^N \langle \nabla U(\bar{V}^{i,N}_{\kappa_\gamma(s)}) - \nabla U(\bm{V}_{s}), e^{i,N}_s \rangle \leq C(1+ d_\theta/N + d_x)^2\gamma -\frac{\mu}{2}\Vert e_s\Vert^2,
  \end{align}
  at which point substituting this into \eqref{eq: newdif}, the result follows in the same manner as the conclusion to the proof of Proposition \ref{prop: discr error 1}.
  \end{proof}
  
  \section{Proofs of Section~\ref{sec: stochastic gradient}}
  
  In this section we prove Theorem \ref{thm:main stoch}. To do so let us first define analogous versions of \eqref{eq: Y defn} for \eqref{eq: theta stoch},\eqref{eq: X stoch}, specifically
  \begin{equation*}
  \tilde{V}^{i,N}_n:= (\tilde{\theta}_n, \tilde{X}^{i,N}_n), \;\;\;\tilde{Z}^N_n:= (\tilde{\theta}_n, N^{-1/2}\tilde{X}^{1,N}_n, ..., N^{-1/2}\tilde{X}^{N,N}_n).
  \end{equation*}
  Furthermore, as in \eqref{eq: interpolation} we may define the stochastic interpolation of \eqref{eq: theta stoch},\eqref{eq: X stoch} as
  \begin{align}\label{eq: interpolation stoch}
  \hat{\theta}_{t} &= \theta_0-\frac{1}{N}\sum_{j=1}^N\int^{t}_0  h_\theta(\hat{\theta}_{\kappa_\gamma(s)}\hat{X}_{\kappa_\gamma(s)}^{j, N}, q_{\kappa_\gamma(s)/\gamma+1}) d s+\sqrt{\frac{2}{N}}\int^{t}_0 d\bm{B}^{0,N}_s, \\
  \hat{X}_t^{i, N} &= X_0^{i, N} - \int^{t}_0 h_x(\hat{\theta}_{\kappa_\gamma(s)}, \hat{X}_{\kappa_\gamma(s)}^{i, N}, q_{\kappa_\gamma(s)/\gamma+1}) d s + \sqrt{2}\int^{t}_0 d\bm{B}_s^{i, N}.
  \end{align}
  so that $\tilde{\theta}_n=\hat{\theta}_{n\gamma}$ (and $\hat{\theta}_{\kappa_\gamma(t)}=\tilde{\theta}_{\kappa_\gamma(t)/\gamma}$), and likewise for all other processes. Note that  $q_{n+1}$ is independent of $\sigma(\tilde{Z}^N_0,\tilde{Z}^N_{1},..., \tilde{Z}^N_{n})$, but not independent of $\tilde{Z}^N_m$ for $m>n$. Furthermore, in keeping with prior conventions we define
  \begin{equation}
  \hat{V}^{i,N}_n:= (\hat{\theta}_n, \hat{X}^{i,N}_n), \;\;\;\hat{Z}^{i,N}_n:= (\hat{\theta}_n, N^{-1/2}\hat{X}^{1,N}_n, ..., N^{-1/2}\hat{X}^{N,N}_n).
  \end{equation}
  Observe that the identities \eqref{eq: V/Z identity} continue to hold for these new processes. One notes that Lemmas \ref{lem: numerics bdd stoch} and \ref{lem: incr bdd stoch} and Proposition \ref{prop: stoch discr error} are very similar in statement and proof to Lemmas \ref{lem: numerics bdd moment}, \ref{lem: numerics bdds} and Proposition \ref{prop: discr error 1}, adjusting for the stochastic gradient by replacing $\nabla U(V^{i,N}_n)$ with $h(\tilde{V}^{i,N}_n, q_{n+1})$ (and likewise $\nabla_x U$ with $h_x$ and $\nabla_\theta U$ with $h_\theta$).
  
  \begin{lmm}\label{lem: numerics bdd stoch}
  Let \Cref{Lipschitz assump},~\Cref{Conv assump}, \Cref{Moments assump} and \Cref{Stoch Gradient assump} hold. Then for every $\gamma_0\in(0,\frac{\mu}{2m})$ there exists a constant a constant $C_3>0$ independent of $t,n,N,\gamma, d_\theta, d_x$ such that for every $\gamma \in (0, \gamma_0)$, $n \in \mathbb{N}$, $t,s \geq 0$ and $\lvert t-s\rvert \leq \gamma$
  \begin{equation}\label{eq: stoch moment bound}
      \mathbb{E}[\lvert \lvert \tilde{Z}^N_n \rvert \rvert^2]\leq C(1+d_\theta/N+d_x).
  \end{equation}
  \end{lmm}
  \begin{proof}
      We follow the proof of Lemma \ref{lem: numerics bdd moment}. Specifically one has from \eqref{eq: moment bound} that
      \begin{align}\label{eq: moment bound stoch}
  \mathbb{E}\left[\Vert \tilde{Z}^N_{n+1} -z^\star\Vert^2 \right]&\leq  \mathbb{E}\left[\Vert \tilde{Z}^N_n -z^\star\Vert^2\right]- \frac{2\gamma}{N}\mathbb{E}\left[\sum_{i=1}^N \langle \tilde{V}^{i,N}_n-v^\star,h(\tilde{V}^{i,N}_n, q_{n+1}) -\nabla U(v^\star)\rangle\right] \nonumber \\
  &+ 2\gamma(\frac{d_\theta}{N}+ d_x)+\frac{\gamma^2}{N} \mathbb{E}\left[\sum_{i=1}^N \Vert h(\tilde{V}^{i,N}_n, q_{n+1}) -\nabla U(v^\star)\Vert ^2\right] .
  \end{align}
  Note that $q_{n+1}$ is independent of $\tilde{Z}^N_n$, and therefore of $\tilde{V}^{i,N}_n$ for $i=1,2,...,N$. As a result, by \Cref{Stoch Gradient assump} one has
  \begin{equation*}
  \mathbb{E}\left[\sum_{i=1}^N \langle \tilde{V}^{i,N}_n-v^\star,h(\tilde{V}^{i,N}_n, q_{n+1}) -\nabla U(v^\star)\rangle\right] = \sum_{i=1}^N \mathbb{E}\left[\langle \tilde{V}^{i,N}_n-v^\star,\nabla U(\tilde{V}^{i,N}_n) -\nabla U(v^\star)\rangle\right].
  \end{equation*}
  Therefore, applying the convexity assumption \Cref{Conv assump}, \Cref{Stoch Gradient assump} and the triangle inequality one has by \eqref{eq: V/Z identity} that
      \begin{align*}
  \mathbb{E}\left[\Vert \tilde{Z}^N_{n+1} -z^\star\Vert^2 \right]&\leq  \mathbb{E}\left[\Vert \tilde{Z}^N_n -z^\star\Vert^2\right]- 2\gamma \mu \sum_{i=1}^N\mathbb{E}\left[ \Vert \tilde{V}^{i,N}_n-z^\star \Vert ^2\right] \\
  &+ 2\gamma(\frac{d_\theta}{N}+ d_x)+\frac{\gamma^2}{N} \left[\sum_{i=1}^N 2m(1+\mathbb{E} \Vert \tilde{V}^{i,N}_n \Vert^2 )\right] +2\gamma^2 \Vert \nabla U(v^\star)\Vert^2  \\
  &\leq  \mathbb{E}\left[\Vert \tilde{Z}^N_n -z^\star\Vert^2\right]- 2\gamma \mu \mathbb{E}\left[ \Vert \tilde{Z}^N_n-z^\star \Vert ^2\right]  \\
  &+ 2\gamma(\frac{d_\theta}{N}+ d_x)+2\gamma^2 m(1+\mathbb{E} \Vert \tilde{Z}^N_n \Vert^2 )+2\gamma^2 \Vert \nabla U(v^\star)\Vert^2  \\
  &\leq  (1-2\gamma\mu+4\gamma^2 m)\mathbb{E}\left[\Vert \tilde{Z}^N_n -z^\star\Vert^2\right] + 2\gamma(\frac{d_\theta}{N}+ d_x)+2\gamma^2 (m+2 m\Vert z^\star\Vert^2 +\Vert \nabla U(v^\star)\Vert^2 ). 
  \end{align*}
  Now we calculate for which values of $\gamma$ one has $\lvert 1-2\gamma\mu+4\gamma^2 m  \rvert<1$. 
  %
  By standard calculus one observes that $1-2\gamma\mu+4\gamma^2 m<1$ for $\gamma \in (0,\frac{\mu}{2m})$, and furthermore the minimum point of $\gamma \mapsto 1-2\gamma\mu+4\gamma^2 m$ is $1-\frac{\mu^2}{4m}$, so that since $\frac{\mu^2}{4m} <1/4 $ by Remark \ref{remark: restriction on m}, one can set $\gamma_0 \in (0,\frac{\mu}{2m})$. Now given that $\lvert 1-2\gamma\mu+4\gamma^2 m  \rvert<1$, one may proceed as in the proof of Lemma \ref{lem: numerics bdd moment} to conclude the result.
  \end{proof}
  \begin{lmm}\label{lem: incr bdd stoch}
  Let \Cref{Lipschitz assump},~\Cref{Conv assump}, \Cref{Moments assump} and \Cref{Stoch Gradient assump} hold and consider the interpolated discretisation \eqref{eq: interpolation stoch}. Then for every $\gamma_0\in(0,\frac{\mu}{2m})$ there exists a constant $C>0$ independent of $t,n,N,\gamma, d_\theta, d_x$ such that for every $\gamma \in (0, \gamma_0)$, $n \in \mathbb{N}$, $t\geq 0$
  \begin{equation}\label{eq: lem bound stoch 2}
      \mathbb{E}[\lvert \lvert \hat{Z}^N_t-\hat{Z}^N_{\kappa_\gamma(t)} \rvert \rvert^2]\leq C(1+d_\theta/N+d_x)\gamma.
  \end{equation}
  \end{lmm}
  \begin{proof}
  We follow the proof of Lemma \ref{lem: incr bdd stoch}. Using the interpolation $\hat{Z}^N_t$ of $\tilde{Z}^N_n$ given in \eqref{eq: interpolation stoch}, and given that $\hat{Z}^N_{\kappa_\gamma(t)} = \tilde{Z}^N_{\kappa_\gamma(t)/\gamma}$, and since $q_{n+1}$ is independent of $\tilde{V}^{i,N}_n$ for $i=1,2,...,N$, one calculates 
  \begin{align*}
  \mathbb{E}\left[\Vert \hat{Z}^N_t -\hat{Z}^N_{\kappa_\gamma(t)}  \Vert^2 \right]&\leq   2\lvert t-\kappa_\gamma(t) \rvert(\frac{d_\theta}{N}+d_x)+\frac{\lvert t-\kappa_\gamma(t) \rvert^2}{N}  \mathbb{E}\left[\sum_{i=1}^N \Vert h(\tilde{V}^{i,N}_{\kappa_\gamma(t)/\gamma}, q_{\kappa_\gamma(t)/\gamma+1}) 
  \Vert ^2\right] \nonumber \\
  &\leq   2\lvert t-\kappa_\gamma(t) \rvert(\frac{d_\theta}{N}+d_x)+m\lvert t-\kappa_\gamma(t) \rvert^2  \biggr (1+\frac{1}{N}\sum_{i=1}^N \mathbb{E}\left[ \Vert \tilde{V}^{i,N}_{\kappa_\gamma(t)/\gamma} 
  \Vert ^2\right] \biggr),
  \end{align*}
  so that applying Cauchy-Schwartz, Young's inequality the Lipschitz assumption \Cref{Lipschitz assump} and \eqref{eq: V/Z identity} one has
  \begin{align*}
  \mathbb{E}\left[\Vert \hat{Z}^N_t -\hat{Z}^N_{\kappa_\gamma(t)} \Vert^2 \right]&\leq  C\lvert t-\kappa_\gamma(t) \rvert(\frac{d_\theta}{N}+d_x+\frac{1}{N}\sum_{i=1}^N E[\Vert \tilde{V}^{i,N}_{\kappa_\gamma(t)/\gamma} \Vert^2]) \leq C\gamma(\frac{d_\theta}{N}+d_x+\Vert \tilde{Z}^N_{\kappa_\gamma(t)/\gamma} \Vert^2),
  \end{align*}
  and therefore that the stated result follows from Lemma \ref{lem: numerics bdd moment}. 
  \end{proof}
  
  \begin{prpstn} \label{prop: stoch discr error}
  Let \Cref{Lipschitz assump},~\Cref{Conv assump}, \Cref{Moments assump} and \Cref{Stoch Gradient assump} hold. Then for every for every $\gamma_0\in(0,\frac{\mu}{2m})$ there exists a constant $C>0$ independent of $t,n,N,\gamma, d_\theta, d_x$ such that for every $\gamma \in (0, \gamma_0)$ one has
  \begin{equation} \label{eq: Lipschtiz disc error stoch}
  \mathbb{E}\left[\Vert \tilde{\theta}_n -\bm{\theta}^N_{n\gamma} \Vert^2\right]^{1/2} \leq C (1+\sqrt{d_\theta/N+d_x})\gamma^{1/2},
  \end{equation}
  for all $n\in\mathbb{N}$.
  \end{prpstn}
  \begin{proof}
  We follow the proof of Proposition \ref{prop: discr error 1}, with the addition of a new term $\dot{r}_3(t)$ arising from the stochastic gradient. As before we prove
  \begin{equation} \label{eq: Lipschitz disc error sys stoch}
  \mathbb{E}\left[\Vert \hat{Z}^N_t -\bm{Z}^N_{t} \Vert^2\right] \leq C \gamma.
  \end{equation}
  Since $\hat{Z}^N_t$ is an It\^o process, so we can apply It\^o's formula for $x \mapsto \lvert x \rvert^2$ to obtain
  \begin{align*}
  \lvert\lvert \hat{Z}^N_t - \bm{Z}^N_{t} \rvert\rvert^2 = -\frac{2}{N} \sum_{i=1}^N  \int^t_0 \langle h(\hat{V}^{i,N}_{\kappa_\gamma(s)}, q_{\kappa_\gamma(s)/\gamma+1}) - \nabla U(\bm{V}_{s}), \hat{V}^{i,N}_s - \bm{V}^{i,N}_{s} \rangle d s .
  \end{align*}
  For convenience let us define 
  \begin{equation} \label{eq: dif process stoch}
  \dot{e}^{i,N}_t:= \hat{V}^{i,N}_t - \bm{V}^{i,N}_t, \;\; \dot{e}_t = \hat{Z}^N_t - \bm{Z}^N_t, 
  \end{equation}
  so that one has
  \begin{equation}\label{eq: ei identity stoch}
  \frac{1}{N}\sum^N_{i=1} \Vert \dot{e}^{i,N}_t \Vert^2 = \Vert \dot{e}_t \Vert^2. 
  \end{equation}
    Using the above, one may apply It\^o's formula for $x \mapsto e^{\mu t}\lvert x \rvert^2$ to obtain
  \begin{equation}\label{eq: newdifstoch}
   e^{\mu t } \Vert e_t \Vert^2 = \int^t_0 \mu e^{\mu s } \Vert e_s \Vert^2+  2\int^t_0 e^{\mu s }\biggr (- \frac{1}{N} \sum_{i=1}^N \langle h(\hat{V}^{i,N}_{\kappa_\gamma(s)}, q_{\kappa_\gamma(s)/\gamma+1}) - \nabla U(\bm{V}_{s}), \hat{V}^{i,N}_s - \bm{V}^{i,N}_{s} \rangle \biggr) d s.
  \end{equation}
  Then one can add and subtract terms in order to obtain
  \begin{align}\label{eq: dif1 stoch}
 - \frac{1}{N} \sum_{i=1}^N \langle h(\hat{V}^{i,N}_{\kappa_\gamma(s)}, q_{\kappa_\gamma(s)/\gamma+1}) - &\nabla U(\bm{V}_{s}), \hat{V}^{i,N}_s - \bm{V}^{i,N}_{s} \rangle =-\frac{1}{N}\sum_{i=1}^N    \langle \nabla U(\hat{V}^{i,N}_s) - \nabla U(\bm{V}_s^{i,N}), \dot{e}^{i,N}_s \rangle  \nonumber \\ 
  & -\frac{1}{N}\sum_{i=1}^N \langle \nabla U(\hat{V}^{i,N}_{\kappa_\gamma(s)}) - \nabla U(\hat{V}^{i,N}_s), \dot{e}^{i,N}_s \rangle  \nonumber \\
  &-\frac{1}{N}\sum_{i=1}^N    \langle h(\hat{V}^{i,N}_{\kappa_\gamma(s)}, q_{\kappa_\gamma(s)/\gamma+1}) -U(\hat{V}^{i,N}_{\kappa_\gamma(s)}) , \dot{e}^{i,N}_s \rangle  \nonumber \\
  &:=\dot{r}_1(s)+\dot{r}_2(s)+\dot{r}_3(s)
  \end{align}
  For the first term we have by the convexity assumption~\Cref{Conv assump} that
  \begin{align}\label{eq: e1 stoch}
  \mathbb{E}[\dot{r}_1(s)]&:= -\frac{1}{N} \sum_{i=1}^N \mathbb{E}[ \langle \nabla U(\hat{V}^{i,N}_s) - \nabla U(\bm{V}_s), \dot{e}^{i,N}_s \rangle  ]\nonumber \\
  & \leq -\mu \frac{1}{N} \sum_{i=1}^N \mathbb{E}[ \Vert \hat{V}_s^{i, N} - \bm{V}^{i,N}_s\Vert^2 ]  \nonumber \\
  & \leq -\mu \mathbb{E}[ \Vert \dot{e}_s \Vert^2] .
  \end{align}
  For $\dot{r}_2(s)$ the Cauchy-Schwarz inequality, Young's inequality and~\Cref{Lipschitz assump} give
  \begin{align}
  \mathbb{E}[r_2(s)]& := \frac{1}{N}\sum_{i=1}^N\mathbb{E} \langle \nabla U(\hat{V}^{i,N}_s)-\nabla U(\hat{V}^{i,N}_{\kappa_\gamma(s)}) , e^{i,N}_s \rangle ]\nonumber \\
  &\leq \frac{1}{2N}\sum_{i=1}^N  (\frac{1}{2\mu}\mathbb{E}[\Vert \nabla U(\hat{V}^{i,N}_{\kappa_\gamma(s)}) - \nabla U(\hat{V}^{i,N}_s) \rvert\rvert^2 ]+\mathbb{E}[ \frac{\mu}{2}\Vert \dot{e}^{i,N}_s \Vert^2] ) \nonumber \\
  & \leq  C \mathbb{E}[\Vert \hat{Z}^N_{\kappa_\gamma(s)} - \hat{Z}^N_s \Vert^2] +\frac{\mu}{2} \mathbb{E}[\Vert \dot{e}_s\Vert^2 ],
  \end{align}
  so that applying Lemma \ref{lem: incr bdd stoch} one has
\begin{align}\label{eq: e2 stoch}
  \mathbb{E}[r_2(s)]\leq  C(1+d_\theta/N+d_x) \gamma +\frac{\mu}{2} \mathbb{E}[\Vert \dot{e}_s\Vert^2 ],
  \end{align}
  Finally for $\dot{r}_3(s)$ one splits again as
  \begin{align}\label{eq: r3stoch}
  -\frac{1}{N}\sum_{i=1}^N    \langle h(\hat{V}^{i,N}_{\kappa_\gamma(s)}, q_{\kappa_\gamma(s)/\gamma+1}) -&\nabla U(\hat{V}^{i,N}_{\kappa_\gamma(s)}) , \dot{e}^{i,N}_s \rangle =-\frac{1}{N}\sum_{i=1}^N    \langle h(\hat{V}^{i,N}_{\kappa_\gamma(s)}, q_{\kappa_\gamma(s)/\gamma+1}) -\nabla U(\hat{V}^{i,N}_{\kappa_\gamma(s)}) , \dot{e}^{i,N}_ {\kappa_\gamma(s)} \rangle \nonumber \\
  &-\frac{1}{N}\sum_{i=1}^N    \langle h(\hat{V}^{i,N}_{\kappa_\gamma(s)}, q_{\kappa_\gamma(s)/\gamma+1}) -\nabla U(\hat{V}^{i,N}_{\kappa_\gamma(s)}) , \dot{e}^{i,N}_ s-\dot{e}^{i,N}_ {\kappa_\gamma(s)} \rangle \nonumber \\
  &:=\dot{r}_{3,1}(s)+\dot{r}_{3,2}(s).
  \end{align}
  Note that since $q_{\kappa_\gamma(s)/\gamma+1}$ is independent of $\hat{V}^{i,N}_{\kappa_\gamma(s)}= \tilde{V}^{i,N}_{\kappa_\gamma(s)/\gamma}$, by assumption that the gradient is unbiased in \Cref{Stoch Gradient assump}, one has that $\dot{r}_{3,1}(s)=0$. For $\dot{r}_{3,2}(s)$ one calculates (recalling $h=(h_\theta, h_x)$)
  \begin{align*}
  \dot{r}_{3,2}(s)&=-\frac{1}{N}\sum_{i=1}^N \lvert s-\kappa_\gamma(s) \rvert   \biggr\langle h(\hat{V}^{i,N}_{\kappa_\gamma(s)}, q_{\kappa_\gamma(s)/\gamma+1}) -\nabla U(\hat{V}^{i,N}_{\kappa_\gamma(s)}), \nonumber \\
  &\biggr(-\frac{1}{N}\sum_{j=1}^N h_\theta(\hat{V}^{j,N}_{\kappa_\gamma(s)}, q_{\kappa_\gamma(s)/\gamma+1}), -h_x(\hat{V}^{i,N}_{\kappa_\gamma(s)}, q_{\kappa_\gamma(s)/\gamma+1}) \biggr) \biggr \rangle \nonumber \\
  & =\lvert t-\kappa_\gamma(s) \rvert \frac{1}{N}\sum_{i=1}^N \biggr \langle h_\theta(\hat{V}^{i,N}_{\kappa_\gamma(s)}, q_{\kappa_\gamma(s)/\gamma+1})-\nabla _\theta U(\hat{V}^{i,N}_{\kappa_\gamma(s)}), \frac{1}{N}\sum_{j=1}^N h_\theta(\hat{V}^{j,N}_{\kappa_\gamma(s)}, q_{\kappa_\gamma(s)/\gamma+1}) \biggr \rangle \nonumber \\
  &+\lvert t-\kappa_\gamma(s) \rvert \frac{1}{N}\sum_{i=1}^N \langle h_x(\hat{V}^{i,N}_{\kappa_\gamma(s)}, q_{\kappa_\gamma(s)/\gamma+1})-\nabla _x U(\hat{V}^{i,N}_{\kappa_\gamma(s)}), h_x(\hat{V}^{i,N}_{\kappa_\gamma(s)}, q_{\kappa_\gamma(s)/\gamma+1}) \rangle \nonumber \\
  & =\lvert t-\kappa_\gamma(s) \rvert \frac{1}{N^2}\biggr \Vert \sum_{i=1}^N h_\theta(\hat{V}^{j,N}_{\kappa_\gamma(s)}, q_{\kappa_\gamma(s)/\gamma+1}) \biggr \Vert ^2+\lvert s-\kappa_\gamma(s) \rvert \frac{1}{N} \sum_{i=1}^N \Vert h_x(\hat{V}^{i,N}_{\kappa_\gamma(s)}, q_{\kappa_\gamma(s)/\gamma+1})\Vert^2 \nonumber \\
  &- \lvert t-\kappa_\gamma(s) \rvert \biggr \langle \frac{1}{N}\sum_{i=1}^N \nabla _\theta U(\hat{V}^{i,N}_{\kappa_\gamma(s)}), \frac{1}{N}\sum_{j=1}^N h_\theta(\hat{V}^{j,N}_{\kappa_\gamma(s)}, q_{\kappa_\gamma(s)/\gamma+1}) \biggr \rangle \nonumber \\
  & - \lvert t-\kappa_\gamma(s) \rvert  \frac{1}{N}\sum_{i=1}^N \langle \nabla _x U(\hat{V}^{i,N}_{\kappa_\gamma(s)}), h_x(\hat{V}^{i,N}_{\kappa_\gamma(s)}, q_{\kappa_\gamma(s)/\gamma+1}) \rangle 
  \end{align*}
  so that one sees that, since $\hat{V}^{i,N}_{\kappa_\gamma(s)}=\tilde{V}^{i,N}_{\kappa_\gamma(s)/\gamma}$ is independent of $q_{\kappa_\gamma(s)/\gamma+1}$, by \Cref{Stoch Gradient assump} when one applies expectation the second half of the final two inner products become equal to the first, meaning the final two lines are bounded above in expectation by $0$. As a result, one may apply \eqref{eq: sum identity}, \Cref{Stoch Gradient assump} and Lemma \ref{lem: numerics bdd stoch} to obtain
  \begin{align}\label{eq: r32stoch}
  \mathbb{E}[\dot{r}_{3,2}(s) ]&\leq \lvert s-\kappa_\gamma(s) \rvert \frac{1}{N}\sum_{i=1}^N \mathbb{E}[ \Vert h(\hat{V}^{i,N}_{\kappa_\gamma(s)}) \Vert^2] \nonumber \\
  & \leq m\lvert t-\kappa_\gamma(s) \rvert (1+\mathbb{E}[\Vert \hat{Z}^{i,N}_{\kappa_\gamma(s)} \Vert^2] ) \nonumber \\
  & \leq C(1+d_\theta/N+d_x) \gamma
  \end{align}
  so that combining~\eqref{eq: e1 stoch}, ~\eqref{eq: e2 stoch}, ~\eqref{eq: r3stoch} and~\eqref{eq: r32stoch} into~\eqref{eq: dif1 stoch} and taking expectation one obtains
  \begin{equation*}
    - \frac{1}{N} \sum_{i=1}^N \langle h(\hat{V}^{i,N}_{\kappa_\gamma(s)}, q_{\kappa_\gamma(s)/\gamma+1}) - \nabla U(\bm{V}_{s}), \hat{V}^{i,N}_s - \bm{V}^{i,N}_{s} \rangle \leq C(1+d_\theta/N+d_x) - \frac{\mu}{2}\Vert \dot{e}_s \Vert^2
  \end{equation*}
  so that the result follows in the same way as the conclusion of the proof of Proposition \ref{prop: discr error 1}.
  \end{proof}

    \section{Convergence to Wasserstein gradient flow}
  \label{app:gf}

In the following, we explicitly distinguish the $\theta$-component of the solution of the gradient flow ODE~\eqref{eq: Cont EM} and the $\theta$-component of the IPS~\eqref{eq: IPS noised} and denote them by $\bm{\theta}_t$ and $\bm{\theta}^N_t$, respectively.
  
For any $t\geq 0$, we have
\begin{align}
\label{eq:sde_exact}
      \bm{X_t} &= X_0-\int_0^t \nabla_x U(\bm{\theta}_s, \bm{X}_s)d s +\sqrt{2}\bm{B}_t\\
      \bm{\theta_t} &= \theta_0+\int_0^t \left[-\int_{\mathbb{R}^{d_x}}\nabla_\theta U(\bm{\theta}_s, x)\nu_s(x)d x\right]d s \notag,
\end{align}
where $\nu_t:\mathbb{R}^d\to\mathbb{R}$ denotes the density of $\mathcal{L}(X_t)$, and additionally
\begin{align*}
      \bm{X_t}^{j,N} &= X_0-\int_0^t \nabla_x U(\bm{\theta}^N_s, \bm{X}_s^{j,N})d s +\sqrt{2}\bm{B}_t^{j,N}\\
      \bm{\theta_t}^{N} &= \theta_0-\int_0^t \left[\frac{1}{N}\sum_{\ell=1}^N\nabla_\theta U(\bm{\theta}^N_s, \bm{X}_s^{\ell, N})\right]d s + \sqrt{\frac{2}{N}}\bm{B}_t^{0,N},
\end{align*}
for $j=1, \dots, N$.

Similarly to Appendix~\ref{app:ergodicity} we consider the rescaled continuous process
\begin{align*}
  \bm{Z}^N_t &:=(\bm{\theta}^N_t, N^{-1/2}\bm{X}^{1,N}_t, \ldots, N^{-1/2}\bm{X}^{N, N}_t),
\end{align*}
and
\begin{align*}
  \bm{Z}_t &:=(\bm{\theta}_t, N^{-1/2}\widetilde{\bm{X}}^{1,N}_t, \ldots, N^{-1/2}\widetilde{\bm{X}}^{N, N}_t),
\end{align*}
where $\widetilde{\bm{X}}^{i, N}_t$ for $i=1,\dots, N$ are $N$ independent copies of~\eqref{eq:sde_exact} driven by the same $\sqrt{2}\bm{B}_t^{j,N}$ as the particle system, that is
\begin{align*}
      \widetilde{\bm{X}}_t^{j,N} &= X_0-\int_0^t \nabla_x U(\bm{\theta}_s,  \widetilde{\bm{X}}_s^{j,N})d s +\sqrt{2}\bm{B}_t^{j,N}.
\end{align*}
For notational convenience we also define
  \begin{equation*}
  \bm{V}^{i,N}_t=(\bm{\theta}^N_t, \bm{X}^{i,N}_t), \qquad \widetilde{\bm{V}}^{i,N}_t=(\bm{\theta}_t, \widetilde{\bm{X}}^{i,N}_t).
  \end{equation*}
We also recall that
  \begin{equation*}
  \frac{1}{N}\sum^N_{i=1} \Vert \bm{V}^{i,N}_t \Vert^2 = \Vert \bm{Z}^N_t \Vert^2,\;\;\; \frac{1}{N}\sum^N_{i=1} \Vert \widetilde{\bm{V}}^{i,N}_t \Vert^2 = \Vert \bm{Z}_t \Vert^2.
  \end{equation*}
  
The propagation of chaos estimates in~\eqref{eq:poc} follow using a contraction argument combined with the convexity assumption \Cref{Conv assump}  as in \cite[Theorem 3.3]{malrieu2001logarithmic}. 
\begin{proof}[Proof of Proposition~\ref{prop:poc}]

Using It\^o's formula we have
      \begin{align}
      \label{eq:poc1}
  e^{\mu t}\mathbb{E} \left[\|\bm{Z}^N_t - \bm{Z}_t\|^2 \right]&= \mathbb{E}\left[ \int^t_0 \mu e^{\mu s} \|\bm{Z}^N_s - \bm{Z}_s\|^2 ds\right]\nonumber \\
  &- 2 \mathbb{E} \left[ \int_0^t e^{\mu s}\left\langle \bm{\theta}^N_s - \bm{\theta}_s, \frac{1}{N}\sum_{i=1}^N\nabla_\theta U(\bm{V}^{i,N}_s) - \int_{\mathbb{R}^{d_x}}\nabla_\theta U(\bm{\theta}_s, x)\nu_s(x)dx\right\rangle d s \right] \nonumber \\
  &-2\frac{1}{N}\sum_{i=1}^N \mathbb{E}\left[ \int_0^t e^{\mu s} \left\langle \bm{X}^{i,N}_s - \widetilde{\bm{X}}^{i,N}_s, (\nabla _xU(\bm{V}^{i,N}_s) - \nabla_x U(\widetilde{\bm{V}}^{i,N}_s))\right\rangle d s \right] +2 \frac{ d_\theta }{N \mu}(e^{\mu t}-1).\notag
  \end{align}
We can the decompose the angle bracket in the first term as
      \begin{align*}
     & \left\langle \bm{\theta}^N_s - \bm{\theta}_s, \frac{1}{N}\sum_{i=1}^N\nabla_\theta U(\bm{V}^{i,N}_s) - \int_{\mathbb{R}^{d_x}}\nabla_\theta U(\bm{\theta}_s, x)\nu_s(x)dx\right\rangle\\
      &\qquad\qquad= \left\langle \bm{\theta}^N_s - \bm{\theta}_s, \frac{1}{N}\sum_{i=1}^N\left(\nabla_\theta U(\bm{V}^{i,N}_s) - \nabla_\theta U(\widetilde{\bm{V}}_s^{i,N})\right)\right\rangle \\
      &\qquad\qquad+ \left\langle \bm{\theta}^N_s - \bm{\theta}_s, \frac{1}{N}\sum_{i=1}^N\left(\nabla_\theta U(\widetilde{\bm{V}}_s^{i,N})- \int_{\mathbb{R}^{d_x}}\nabla_\theta U(\bm{\theta}_s, x)\nu_s(x)dx\right)\right\rangle.
  \end{align*}

To control the last term in the above we use the Cauchy-Schwarz inequality
\begin{align*}
&- \mathbb{E} \left[  \left\langle \bm{\theta}^N_s - \bm{\theta}_s, \sum_{i=1}^N\left(\nabla_\theta U(\widetilde{\bm{V}}_s^{i,N})- \int_{\mathbb{R}^{d_x}}\nabla_\theta U(\bm{\theta}_s, x)\nu_s(x)dx\right)\right\rangle \right] \\
&\qquad\qquad\leq  \mathbb{E} \left[ \Vert \bm{\theta}^N_s - \bm{\theta}_s\Vert^2\right]^{1/2}\mathbb{E} \left[\left\lVert\sum_{i=1}^N\left(\nabla_\theta U(\widetilde{\bm{V}}_s^{i,N})- \int_{\mathbb{R}^{d_x}}\nabla_\theta U(\bm{\theta}_s, x)\nu_s(x)dx\right) \right\rVert^2\right]^{1/2}.
\end{align*}
Let use denote $\xi_i(s) = \nabla_\theta U(\widetilde{\bm{V}}_s^{i,N})- \int_{\mathbb{R}^{d_x}}\nabla_\theta U(\bm{\theta}_s, x)\nu_s(x)dx$, then
\begin{align*}
\mathbb{E} \left[\left\lVert\sum_{i=1}^N\left(\nabla_\theta U(\widetilde{\bm{V}}_s^{i,N})- \int_{\mathbb{R}^{d_x}}\nabla_\theta U(\bm{\theta}_s, x)\nu_s(x)dx\right) \right\rVert^2\right] &= \sum_{i=1}^N \mathbb{E}[\Vert \xi_i(s)\vert^2]+2\sum_{1\leq i\leq j\leq N}\mathbb{E}[ \xi_i(s)\xi_j(s)]\\
&= \sum_{i=1}^N \mathbb{E}[\Vert \xi_i(s)\Vert^2]
\end{align*}
due to the independence assumption on $\widetilde{\bm{V}}_s^{i,N}$.
Then, using the fact that $\textrm{var}(Y)\leq \mathbb{E}[Y^2]$ for all random variables $Y$ and using the Lipschitz continuity in \Cref{Lipschitz assump}, we have
\begin{align*}
\mathbb{E}[\Vert \xi_i(s)\Vert^2] &= \mathbb{E}[\Vert  \nabla_\theta U(\widetilde{\bm{V}}_s^{i,N})- \int_{\mathbb{R}^{d_x}}\nabla_\theta U(\bm{\theta}_s, x)\nu_s(x)dx\Vert^2]\\
&\leq \mathbb{E}[\Vert  \nabla_\theta U(\widetilde{\bm{V}}_s^{i,N})\Vert^2]\\
&\leq 2L^2 \mathbb{E}[\Vert  \widetilde{\bm{V}}_s^{i,N}-v^*\Vert^2],
\end{align*}
where $v^*$ denotes the minimiser of $U$. The above expression is finite for all $s$ by \cite[Lemma 3.4]{chaintron2021propagation} . 

Let us denote $C:= \sqrt{2}L \sup_{s\geq0}\mathbb{E}[\Vert  \widetilde{\bm{V}}_s^{i,N}-v^*\Vert^2]^{1/2}<\infty$. We can write
      \begin{align*}
    e^{\mu t}\mathbb{E}\left[\|\bm{Z}^N_t - \bm{Z}_t\|^2 \right]&= \mathbb{E}\left[ \int^t_0 \mu e^{\mu s} \|\bm{Z}^N_s - \bm{Z}_s\|^2 ds\right] \\
 & - 2 \mathbb{E} \left[\frac{1}{N}\sum_{i=1}^N\int_0^t e^{\mu s}\left\langle \bm{V}^{i,N}_s - \widetilde{\bm{V}}^{i,N}_s, \nabla U(\bm{V}^{i,N}_s) - \nabla U(\widetilde{\bm{V}}^{i,N}_s)\right\rangle d s \right]+ 2 \frac{d_\theta }{N \mu}(e^{\mu t}-1)\\
  & - 2 \mathbb{E} \left[ \int_0^t e^{\mu s}\left\langle \bm{\theta}^N_s - \bm{\theta}_s, \frac{1}{N}\sum_{i=1}^N\left(\nabla_\theta U(\widetilde{\bm{V}}_s^{i,N})- \int_{\mathbb{R}^{d_x}}\nabla_\theta U(\bm{\theta}_s, x)\nu_s(x)dx\right)\right\rangle d s \right]\\
  &\leq \mathbb{E}\left[ \int^t_0 \mu e^{\mu s} \|\bm{Z}^N_s - \bm{Z}_s\|^2 ds\right] -2 \mu \int_0^t e^{\mu s}\mathbb{E}\left[\|\bm{Z}^N_s - \bm{Z}_s\|^2 \right]ds\\
  & + 2 \frac{ d_\theta}{N\mu}e^{\mu t} +\frac{2C}{N^{1/2}}\int_0^t e^{\mu s}\mathbb{E} \left[ \Vert \bm{\theta}^N_s - \bm{\theta}_s\Vert^2\right]^{1/2}ds\\
  &\leq \mathbb{E}\left[ \int^t_0 \mu e^{\mu s} \|\bm{Z}^N_s - \bm{Z}_s\|^2 ds\right]-2\mu \int_0^t e^{\mu s}\mathbb{E}\left[\|\bm{Z}^N_s - \bm{Z}_s\|^2 \right]ds \\
  &+ 2 \frac{ d_\theta}{N\mu}e^{\mu t}+\frac{2C}{N^{1/2}}\int_0^t e^{\mu s}\mathbb{E} \left[ \|\bm{Z}^N_s - \bm{Z}_s\|^2\right]^{1/2}ds,
  \end{align*}
where we used the convexity assumption \Cref{Conv assump} for the second term. Then applying Young' inequality $2ab \leq a^2+b^2$ with $a=e^{\mu s/2}C/(\sqrt{N\mu})$ and $b= \sqrt{\mu}e^{\mu s/2}\left[ \|\bm{Z}^N_s - \bm{Z}_s\|^2\right]^{1/2}$ one has
      \begin{align*}
  e^{\mu t}\mathbb{E}\left[\|\bm{Z}^N_t - \bm{Z}_t\|^2 \right]&\leq \mathbb{E}\left[ \int^t_0 \mu e^{\mu s} \|\bm{Z}^N_s - \bm{Z}_s\|^2 ds\right]-2\mu \int_0^t e^{\mu s}\mathbb{E}\left[\|\bm{Z}^N_s - \bm{Z}_s\|^2 \right]ds \\
  &+ 2 \frac{ d_\theta}{N\mu}e^{\mu t}+\frac{C^2}{N\mu}\int_0^t e^{\mu s}ds+\int_0^t \mu e^{\mu s}\mathbb{E} \left[ \|\bm{Z}^N_s - \bm{Z}_s\|^2\right]ds \\
  &\leq  \frac{ 2d_\theta +C^2}{N\mu}e^{\mu t},
  \end{align*}
at which point the result follows by dividing through by $e^{\mu t}$.

\end{proof}

\section{Verifying assumptions in the Logistic Regression model}\label{app:verify_a1_a2}
In this section, we detail the convexity and Lipschitzness properties of the example given in Section~\ref{sec:logistic_synth}.
As shown in \cite[Lemma S18]{de2021efficient}, we have
\begin{align*}
    U(\theta, x) =-\log p_\theta(x, y) =  (d_x/2)\log(2\uppi\sigma^2) - \sum_{j=1}^{d_y}\left(y_j\log(s(v_j^Tx))+(1-y_j)\log(s(-v_j^Tx))\right)+\frac{\norm{x-\theta}^2}{2\sigma^2},
\end{align*}
and
\begin{align*}
    \nabla_\theta  U(\theta, x) &= -\frac{ x-\theta }{\sigma^2},\qquad\qquad
    \nabla_x U(\theta, x) = \frac{x-\theta}{\sigma^2} - \sum_{j=1}^{d_y}\left(y_j-s(v_j^Tx)\right)v_j.
\end{align*}
To check that \Cref{Lipschitz assump} holds observe that
\begin{align*}
    \norm{\nabla U(v)-\nabla U(v')} &\leq\norm{\nabla_\theta  U(\theta, x)-\nabla_\theta  U(\theta', x')}+ \norm{\nabla_x  U(\theta, x)-\nabla_x  U(\theta', x')}\\
    &\leq \frac{2}{\sigma^2}(\norm{x-x'} + \norm{\theta-\theta'})+ \sum_{j=1}^{d_y} \vert s(v_j^Tx)-s(v_j^Tx')\vert \norm{ v_j}\\
    &\leq \frac{2}{\sigma^2}(\norm{x-x'} + \norm{\theta-\theta'})+ \frac{1}{4}\norm{x-x'}\sum_{j=1}^{d_y} \norm{ v_j}^2\\ 
 &\leq (2\sigma^{-2}+1/4\sum_{j=1}^{d_y} \norm{ v_j}^2)\left(\norm{\theta-\theta'}+\norm{ x'-x}\right),
\end{align*}
where we used the fact that the logistic function is Lipschitz continuous with Lipschitz constant $\sup_{u\in \mathbb{R}} \vert s'(u)\vert = 1/4$. 

For \Cref{Conv assump}, we can only check strict convexity, and not directly \Cref{Conv assump}. Consider the Hessian of $U$
\begin{align*}
 \nabla^2 U(v) &= \frac{1}{\sigma^2}\begin{pmatrix}
 \textsf{Id}_{d_\theta}&-\textsf{Id}_{d_x}\\
 -\textsf{Id}_{d_x}&\textsf{Id}_{d_\theta}
 \end{pmatrix} + \sum_{j=1}^{d_y} s(v_j^Tx)(1-s(v_j^Tx))v_j\otimes v_j.
\end{align*}
The leftmost matrix has $2d_\theta$ strictly positive eigenvalues, and, as shown in \cite[Proposition 1, Appendix E.2]{kuntz2023particle} the rightmost matrix is positive definite, it follows that $ \nabla^2 U(v)$ is positive definite and thus $U$ is strictly convex.
\end{appendix}

\end{document}